\documentclass[aps,prb,superscriptaddress,twocolumn,floatfix]{revtex4-1}
\usepackage{graphicx}
\usepackage{graphics}
\usepackage{amsmath}
\usepackage{amsfonts}
\usepackage{subcaption}
\usepackage{bm}
\usepackage{color}
\usepackage[version=4]{mhchem}
\usepackage{pgf}
\usepackage{siunitx}
\usepackage{tikz}
\usepackage{calculator}
\usepackage{lipsum}
\usepackage{tabularx}
\usepackage{hyperref}
\usepackage{soul}
\usepackage{esvect}
\usepackage{float}
\usetikzlibrary{calc}
\def\vec#1{\boldsymbol{#1}}

\hypersetup{linkcolor=red,colorlinks=true}

\begin{document}

\title{Theoretical investigation of twin boundaries in WO$_3$: Structure, properties and implications for superconductivity}

\author{No\'{e} Mascello}
\affiliation{Materials Theory, ETH Zurich, Wolfgang-Pauli-Strasse 27, CH 8093 Zurich, Switzerland}

\author{Nicola A. Spaldin}
\affiliation{Materials Theory, ETH Zurich, Wolfgang-Pauli-Strasse 27, CH 8093 Zurich, Switzerland}

\author{Awadhesh Narayan}
\email{awadhesh@iisc.ac.in}
\affiliation{Materials Theory, ETH Zurich, Wolfgang-Pauli-Strasse 27, CH 8093 Zurich, Switzerland}
\affiliation{Solid State and Structural Chemistry Unit, Indian Institute of Science, IN 560012 Bangalore, India}

\author{Quintin N. Meier}
\email{quintin.meier@mat.ethz.ch}
\affiliation{Materials Theory, ETH Zurich, Wolfgang-Pauli-Strasse 27, CH 8093 Zurich, Switzerland}
\email{quintin.meier@mat.ethz.ch}

\date{\today}

\begin{abstract}
 We present a theoretical study of the structure and functionality of ferroelastic domain walls in tungsten trioxide, \ce{WO3}. \ce{WO3} has a rich structural phase diagram, with the stability and properties of the various structural phases strongly affected both by temperature and by electron doping. The existence of superconductivity is of particular interest, with the underlying mechanism as of now not well understood. In addition, reports of enhanced superconductivity at structural domain walls are particularly intriguing. Focusing specifically on the orthorhombic $\beta$ phase, we calculate the structure and properties of the domain walls both with and without electron doping. We use two theoretical approaches: Landau-Ginzburg theory, with free energies constructed from symmetry considerations and parameters extracted from our first-principles density functional calculations, and direct calculation using large-scale, GPU-enabled density functional theory. We find that the structure of the $\beta$-phase domain walls resembles that of the bulk tetragonal $\alpha_1$ phase, and that the electronic charge tends to accumulate at the walls. Motivated by this finding, we perform \textit{ab initio} computations of electron-phonon coupling in the bulk $\alpha_1$ structure and extract the superconducting critical temperatures , $T_c$, within Bardeen-Cooper-Schrieffer theory. Our results provide insight into the  experimentally observed unusual trend  of decreasing $T_c$ with increasing electronic charge carrier concentration.
\end{abstract}

\maketitle
\section{Structure and properties of WO$_3$}
\subsection{Introduction}
\begin{figure}[ht]
\begin{tikzpicture}
    \node at (0,0) {\includegraphics[scale=0.35,trim=300 50 300 50, clip]{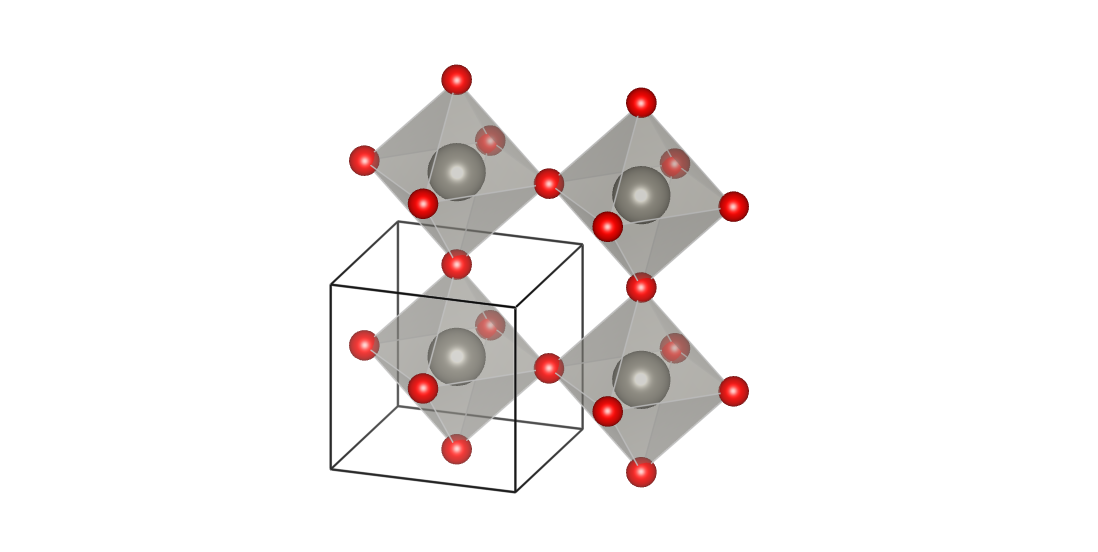}};
    
    \node[circle,outer color=black!65!white,inner color=white,minimum size=0.75cm] (W) at (3,1.5) {};
    \node[circle,outer color=black!15!red,inner color=white!70!red,minimum size=0.25cm] (O) at (3,0) {};
    
    \node[right] at ($(W) + (0.3,0)$) {W};
    \node[right] at ($(O) + (0.3,0)$) {O};
    
    \def \l {1.5}
    \def \x {-3}
    \def \y {-2.8}
    \draw [->,thick] (\x,\y) --++ (-7.5:\l) node[below] {$\left[010\right]$};
    \draw [->,thick] (\x,\y) --++ (-135:\l*0.8) node[below right] {$\left[100\right]$};
    \draw [->,thick] (\x,\y) --++ (90:\l) node[left]
    {$\left[001\right]$};

\end{tikzpicture}
\caption{High-symmetry cubic pseudo-perovskite $Pm\overline{3}m$ structure of \ce{WO3}. The perovskite B sites are occupied by W atoms (grey) which are encapsulated by O atom (red) octahedra. The A sites on the unit cell corners are vacant.}
\label{fig:WO3_Pm3m}
\end{figure}
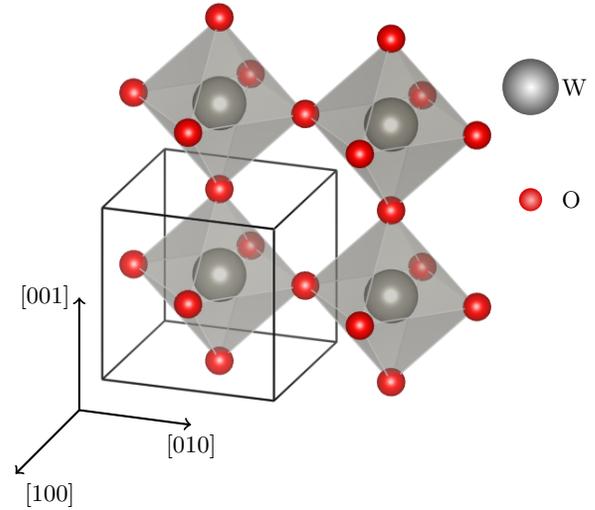

Tungsten trioxide, \ce{WO3}, is a functionally versatile material with possible applications based on electrochromism (smart windows), gasochromism (gas sensors) and photocatalysis \cite{Granqvist2000, Zheng2011, Santato2001}. The high-symmetry structure of \ce{WO3} is that of a perovskite with a vacant A site (see Fig. \ref{fig:WO3_Pm3m}), and it exhibits a series of lower symmetry phases at lower temperature. Both its structure and its properties depend on and can be tuned by doping \cite{Hagenmuller1973}. For instance, while pure \ce{WO3} is an insulator, it becomes metallic upon occupation of the A sites with alkali metal ions. In addition, superconductivity was reported as early as 1964 for Na-doped \ce{WO3}, and \ce{M_xWO_{3-x}} systems are now well established as superconductors (with M usually an alkali metal) \cite{Raub1964, Sweedler1965, Sweedler1965_2, Hubble1971, Shanks1974, Skokan1979, Cadwell1981, Garif'yanov1996, Salje1997, Brusetti2002, Reich2009, Bocarsly2013, Haldolaarachchige2014}. The reported superconducting critical temperatures ($T_c$) for bulk \ce{M_{x}WO_{3-x}} systems are generally less than \SI{2}{\kelvin} \cite{Shanks1974, Garif'yanov1996}, and, interestingly, tend to  decrease with increasing doping above the lowest doping level at which superconductivity is observed \cite{Shanks1974, Haldo2014, Pellegrini2019}. Moreover, different dopants result in superconductivity in different structural phases. More recently, non-bulk high-temperature superconductivity was reported on the surface of dopant-rich islands in \ce{Na_{x}WO_{3-x}} \cite{Reich1999}. Around the same time, superconductivity was discovered in reduced \ce{WO_{3-x}} with a superconducting critical temperature of around \SI{3}{\kelvin} \cite{Aird1998, Aird1998_2, Aird2000}. In this case, the bulk sample was not superconducting, but sheet superconductivity occurred along the ferroelastic domain walls of the reduced \ce{WO_{3-x}} crystals. 

\subsection{Structural phase transitions and domain walls}
\begin{figure}
    \hspace*{-0cm}%
    \begin{tikzpicture}
        \node[] (Pm3m) at (0,3) {\large $Pm\overline{3}m$};
        \node[] (P4nmm) at (0,1.5) {\large $P4/nmm \ (\alpha_1)$};
        \node[] (P4ncc) at (0,0) {\large $P4/ncc \ (\alpha_2)$};
        \node[] (Pbcn) at (0,-1.5) {\large $Pbcn \ (\beta)$};
        \node[] (P21n) at (0, -3) {\large $P2_1/n \ (\gamma)$};
        
        \node[text width=10, align=left] at (1.6, 1.5) {\textcolor{blue}{$M_2^-$}};
        \node[text width=10, align=left] at (1.6, 0) {\textcolor{blue}{$M_2^-, R_5^-$}};
        \node[text width=10, align=left] at (1.6, -1.5) {\textcolor{blue}{$M_2^-, R_5^-, X_5^+, M_2^+$}};
        \node[text width=10, align=left] at (1.6, -3.0) {\textcolor{blue}{$M_2^-, R_5^-, X_5^+, M_2^+$}};
        \draw[blue,->, thick] (Pm3m) -- (P4nmm) node [midway, right] {$M_2^-$};
        \draw[red,->, thick] ($(P4ncc)+(-0.3,0.25)$) -- ($(P4nmm)+(-0.3,-0.25)$) node [midway, left] {$\Gamma_1^+$ ($R$)};
        \draw[red,->, thick] (P4ncc) -- (Pbcn) node [midway, right] {$M_1$ ($Q$)};
        \draw[black!30!green,<-, thick] ($(P4ncc)+(0.3,0.25)$) -- ($(P4nmm)+(0.3,-0.25)$)  node[midway, right] {$Z_3^+$} (P4ncc);
        \draw[black,->, thick] ($(Pbcn)+(0,-0.25)$) -- ($(P21n)+(0,0.25)$)  node[midway, right] {$\Gamma_2^+$} (P4ncc);
        \draw[black, ->, thick] (-2,3) -- (-2,-3) node [midway,left] {\large $T \downarrow$};
        \draw[black, <-, thick] (4.2,3) -- (4.2,-3) node [midway,right] {\large $\rho \uparrow$};
        
    \end{tikzpicture}
    \caption{Symmetries of, and transitions between, the structural phases of \ce{WO3} as a function of decreasing temperature $T$ or increasing amount of electron doping $\rho$.
    The arrows indicate the transition modes using the notation of the structure at the arrow origin. That is blue modes denote modes of the cubic, green of the $\alpha_1$, red of the $\alpha_2$ and black of the $\beta$ structures. Listed in blue on the right are all modes of the cubic structure present in the respective lower-symmetry structures. The $Q$ and $R$ labels at the transitions represented by red arrows denote parameters of the energy expansions that will be introduced later.}
    \label{fig:phases_modes}
\end{figure}
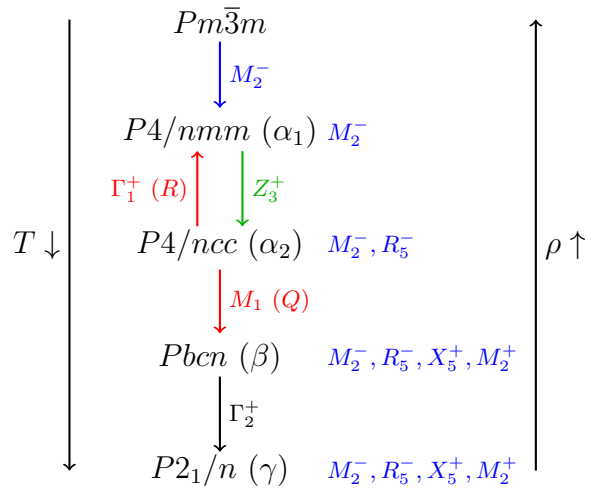

\begin{figure}
    \begin{subfigure}{0.235\textwidth}
        \centering
        \includegraphics[width=\textwidth, trim={58cm 12cm 58cm 12cm},clip]{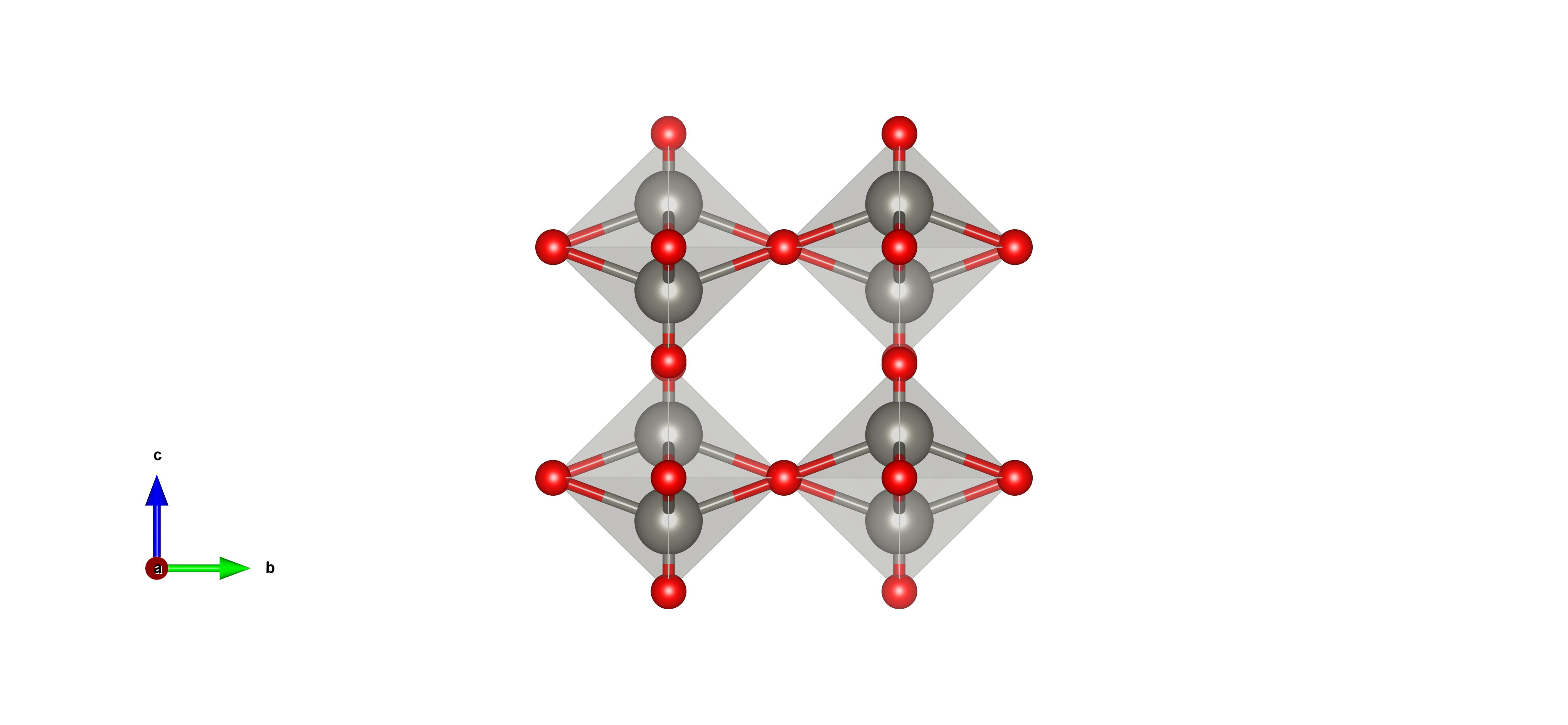}
        \subcaption{$M_2^-$}
    \end{subfigure}
    \begin{subfigure}{0.235\textwidth}
        \centering
        \includegraphics[width=\textwidth, trim={58cm 12cm 58cm 12cm},clip]{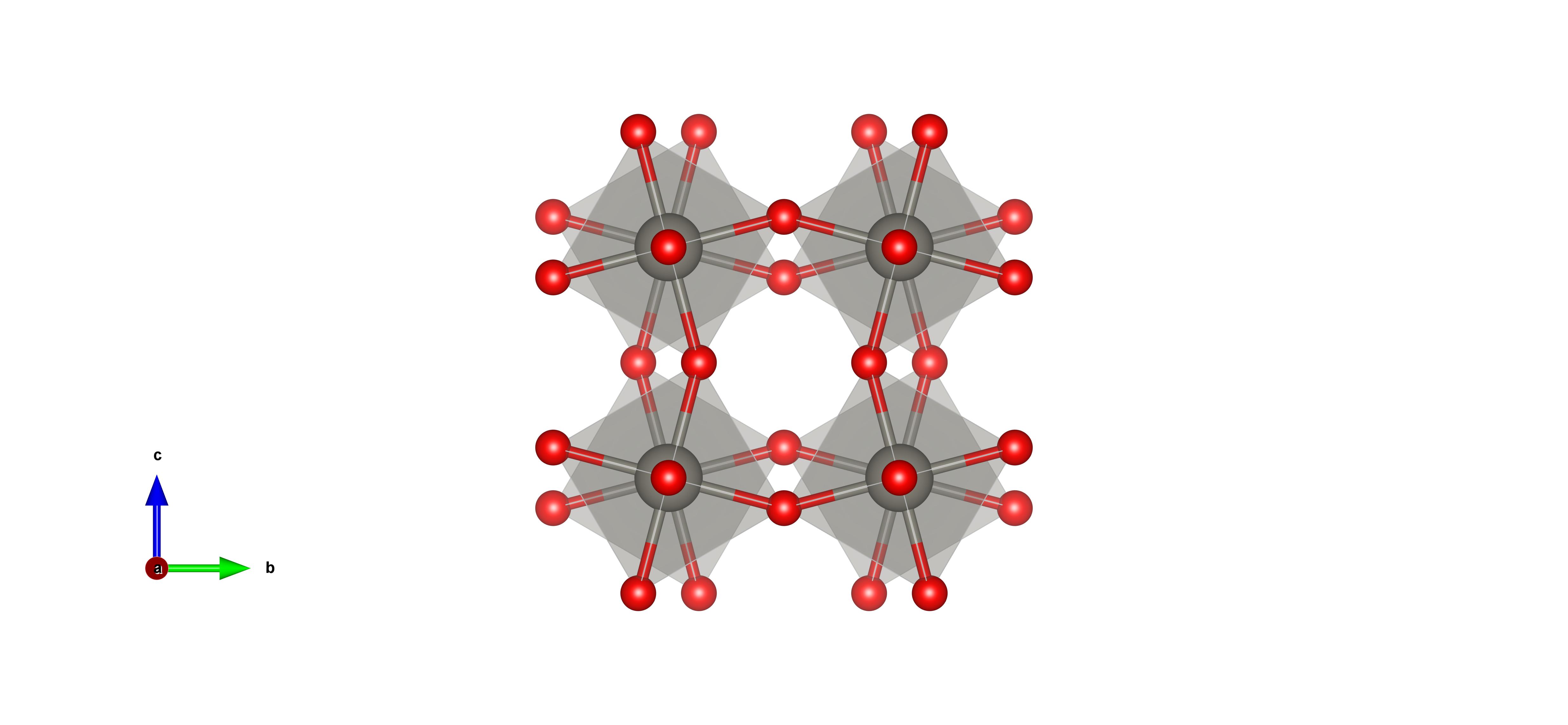}
        \subcaption{$R_5^-$}
    \end{subfigure}
    
    \begin{subfigure}{0.235\textwidth}
        \centering
        \includegraphics[width=\textwidth, trim={58cm 12cm 58cm 12cm},clip]{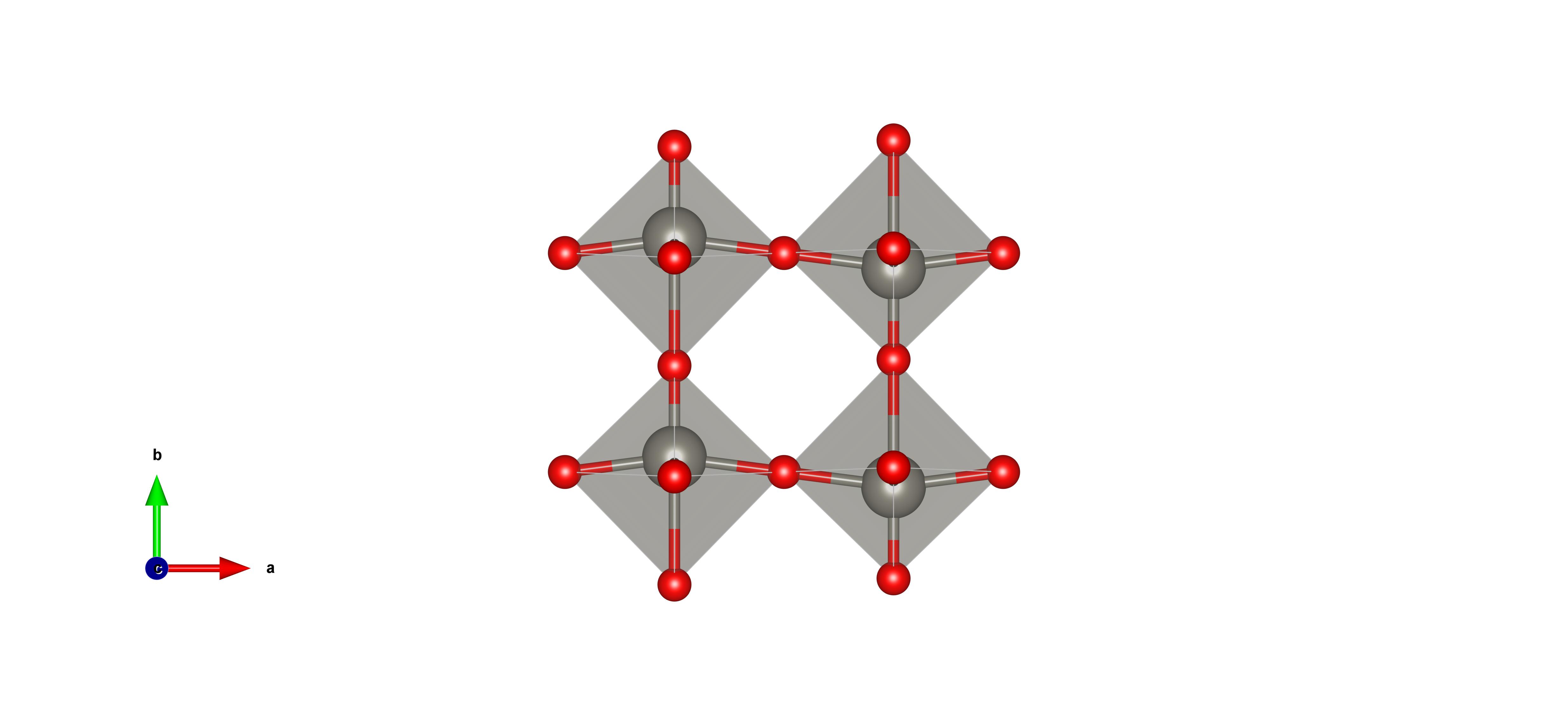}
        \subcaption{$X_5^+$}
    \end{subfigure}
    \begin{subfigure}{0.235\textwidth}
        \centering
        \includegraphics[width=\textwidth, trim={58cm 12cm 58cm 12cm},clip]{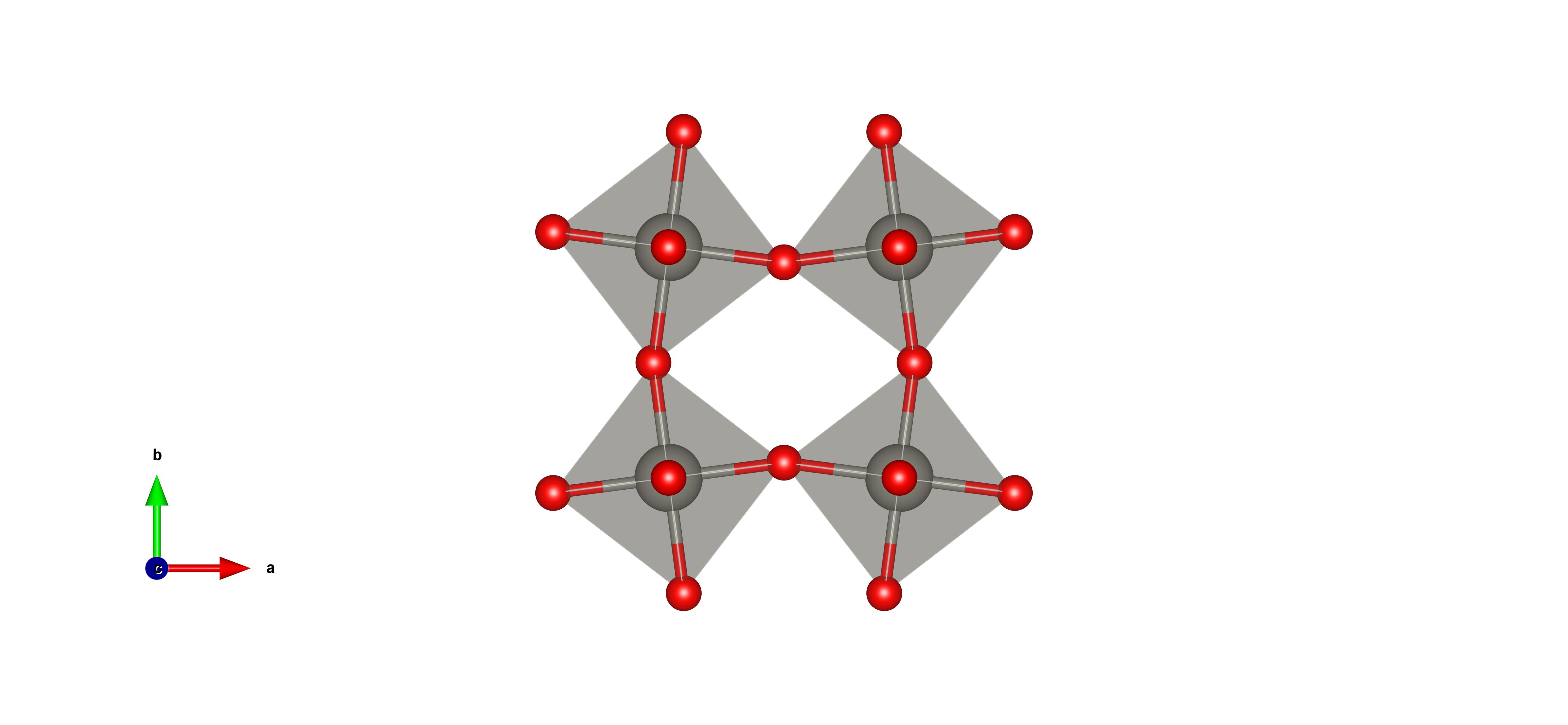}
        \subcaption{$M_2^+$}
    \end{subfigure}
    \caption{Distortions of the cubic \ce{WO3} structure, labelled with the irreducible representations of the cubic unit cell, that lead to the tetragonal $\alpha_1$ and $\alpha_2$, and orthorhombic $\beta$ phases.}
    \label{fig:cubic_distortions}
\end{figure}

At room temperature, \ce{WO3} shows two types of ferroelastic domain walls which correspond to two successive ferroelastic transitions at higher temperatures. Fig. \ref{fig:phases_modes} shows the sequence of structural phases of \ce{WO3} as a function of temperature and doping \cite{Aird1998, Locherer1998, Salje2004, Kim2010}. The high-symmetry cubic $Pm\overline{3}m$ reference structure of \ce{WO3} does not form under standard conditions as \ce{WO3} sublimes before reaching it \cite{Vogt1999, Crichton2003}. Therefore there are no ferroelastic domain walls resulting from a cubic $Pm\overline{3}m$ to tetragonal $P4/nmm$ ($\alpha_1$) transition. Additionally, the subsequent $\alpha_1 - \alpha_2$ transition does not form domain walls since it does not change the point symmetry \cite{Locherer1999_2, Howard2001}. \newline
As temperature is further reduced, the next structural phase transition, and the first ferroelastic transition, is from the tetragonal $\alpha_{2}$ (space group $P4/ncc$) to the orthorhombic $\beta$  phase ($Pbcn$). In terms of distortions from the cubic phase, the tetragonal $\alpha_2$ phase is characterized by two normal modes of the cubic perovskite structure with representations (and wave vectors) $M_2^-$ ($\frac{1}{2},\frac{1}{2},0$) and $R_5^-$ ($\frac{1}{2},\frac{1}{2},\frac{1}{2}$), respectively (see Fig. \ref{fig:phases_modes} and \ref{fig:cubic_distortions}), we refer to these as the cubic $M_2^-$ and the cubic $R_5^-$ modes below. The cubic $M_2^-$ mode (Fig.~\ref{fig:cubic_distortions} (a)) consists of antipolar displacements of the W atoms, and the cubic $R_5^-$ mode (Fig.~\ref{fig:cubic_distortions} (b)) of out-of-phase rotations of the O octahedra with $a^0a^0b^-$ Glazer notation.
The transition from the $\alpha_2$ to the $\beta$ phase then introduces in addition mainly the cubic $X_5^+$ ($\frac{1}{2},0,0$) and the cubic $M_2^+$ ($\frac{1}{2},\frac{1}{2},0$) modes and reorients the already present cubic $M_2^-$ mode from along the $a$ axis to the $(a,b)$ diagonal spatial direction. The $X_5^+$ mode causes further antipolar displacements of the W atoms and the $M_2^+$  mode introduces in-phase rotations of the O octahedra (Glazer notation $a^0a^0b^+$). The primary order parameter with respect to the tetragonal $\alpha_2$ phase is an $M_1$ ($\frac{1}{2},\frac{1}{2},0$) mode, which causes a doubling of the unit cell along the rotation axis, along with antipolar displacements. We refer to the domain walls that form between different orientations of the $\beta$ phase at this transition as $\beta$ domain walls.

Upon further reduction of temperature, the second ferroelastic transition corresponds to the transformation from the orthorhombic $\beta$ phase to the monoclinic $\gamma$ phase ($P2_1/n$). In addition to the monoclinic distortion of the lattice, this transition introduces additional directional components to the already present cubic $X_5^+$ and $R_5^-$ modes, however the additional amplitude of the $X_5^+$ is very small. These changes require no further doubling of the unit cell, so the order parameter of this second transition is an orthorhombic $\Gamma$ mode: $\Gamma_2^+$ ($0,0,0$). We refer to the domain walls that form between different orientations of the $\gamma$ phase at this transition as $\gamma$ domain walls.

\begin{figure}
    \centering
    \begin{tikzpicture}
        \def \l {2}  
        \def \d {0.25} 
        \def \nb {3} 
        \def \nd {35} 
        
        \SQUAREROOT{2}{\sqrttwo}
        \DIVIDE{\l}{\sqrttwo}{\lb}
        
        \begin{scope}
        \clip(-\lb,\lb) rectangle (\nb*\lb,5);
        \draw[red, very thick] (-\lb,0) --++ (0, \nd*\d+\lb);
        \foreach \i  in {0,...,\nb}
        {
            \ifodd \i
                \def \angle {45}
            \else
                \def \angle {135}
            \fi
            
            \DIVIDE{\i}{2}{\s}
            \FLOOR{\s}{\s}

            \foreach \j in {0,...,\nd}
            {
                \draw[blue] (2*\s*\lb,\j*\d) --++ (\angle:\l);
            }
            \draw[red, very thick] (\i*\lb, 0) --++ (0,\nd*\d+\lb);
            
            \node[red] at (\i*\lb-0.3, 4.7) {\Large \textbf{$\beta$}};
            \node[blue] at (\i*\lb-\l/3, 3.2) {\huge \textbf{$\gamma$}};
        }
        \end{scope}
        
    \def \l {1.5}
    \def \x {-2}
    \def \y {0}
    \draw [->,thick] (\x,\y) --++ (0:\l) node[below] {$\left[110\right]$};
    \draw [->,thick] (\x,\y) --++ (45:\l) node[right] {$\left[100\right]$};
    \draw [->,thick] (\x,\y) --++ (90:\l) node[left]
    {$\left[1\overline{1}0\right]$};
    \end{tikzpicture}
    \caption{Idealized schematic of ferroelastic domain walls in \ce{WO3}. The 2D coordinate system corresponds to the pseudocubic directions. Red and blue lines represent $\beta$ domain walls with $\left(110\right)$ and $\gamma$ domain walls with $\left(100\right)$ crystallographic plane orientation, respectively. Domain sizes and domain wall widths are not representative and $\beta$ domain walls can also be present with $\left(1\overline{1}0\right)$ plane orientation.}
    \label{fig:domains}
\end{figure}
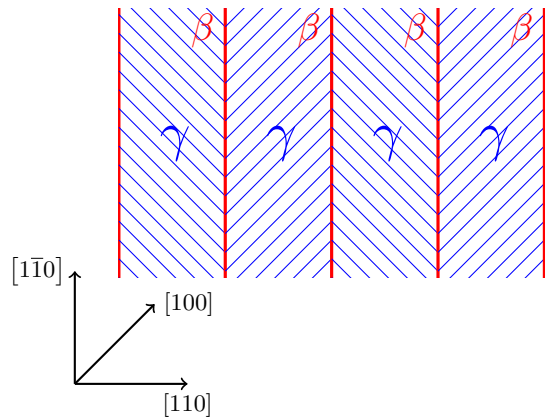

In experiments, the two types of domain walls create a pattern in which the $\gamma$ domain walls (blue in Fig. \ref{fig:domains}) form in a zig-zag manner between the $\beta$ domain walls (red in Fig. \ref{fig:domains}) \cite{Aird2000, Salje2004, Yun2015}. The $\gamma$ domain walls in different $\beta$ domains meet at 90 degree angles, and the $\beta$ and $\gamma$ domain walls are oriented at 45 degrees with respect to each other. Atomic force microscopy (AFM) measurements of epitaxially grown \ce{WO3} films have shown that the two types of domain walls correspond to crystallographic $\left(100\right)$ ($\gamma$) and $\left(110\right)$ ($\beta$) planes \cite{Yun2015}. The same orientations are implied by a strain analysis of the respective ferroelastic transitions \cite{Salje1991, Aizu1970, Sapriel1975}: The planes of vanishing strain for a ferroelastic transition of type 4/mmmFmmm (which is the type of the $\alpha_2-\beta$ transition) correspond to $\left(110\right)$ planes and a transition of type mmF2/m (which the $\beta-\gamma$ transition corresponds to) has $\left(100\right)$ vanishing strain planes.  \newline

\subsection{Questions addressed in this work}
In this work, we investigate theoretically the structure and properties of the $\beta$ domain walls in \ce{WO3}. We start by constructing the Landau-Ginzburg free energy densities using parameters obtained from electronic structure calculations based on density functional theory (DFT), and use these Landau-Ginzburg expressions to calculate the structure and properties of the walls with and without doping. We benchmark our model calculations by also calculating the structure of the $\beta$ domain wall directly from first-principles using a large supercell. Our calculations allow us to extract both the structural and electrostatic changes associated with domain wall formation. We find that the structure at the $\beta$ domain wall resembles that of the bulk tetragonal $\alpha_1$ phase, and that the electrostatic changes cause a local accumulation of electronic charge at the wall. Motivated by these results, we calculate the doping dependence of the critical temperature for electron-phonon mediated superconductivity in the bulk tetragonal $\alpha_1$ structure, and find that Bardeen-Cooper-Schrieffer (BCS) theory reproduces the experimentally observed decrease in $T_c$ with increasing doping in the $\alpha_1$ phase.

\section{Landau-Ginzburg theory of the tetragonal to orthorhombic ($\alpha_2 \to \beta$) transition}

We begin our treatment of the $\beta$ ferroelastic domain walls in \ce{WO3} by constructing the equations describing their free energy density according to the Landau-Ginzburg theory of the $\alpha_2 - \beta$ transition. We then determine the lowest energy domain wall profiles by numerical minimization, and partly by analytical solution, of the free energy density. The detailed analytical solution is presented in appendix section \ref{subsec:analytical_solutions}.

As outlined above, the tetragonal $\alpha_2$ to orthorhombic $\beta$ transition, which occurs at around \SI{1000}{\kelvin}, is driven by the condensation of a single mode with $M_1$ symmetry of the tetragonal phase (see Fig. \ref{fig:phases_modes} and Fig. \ref{fig:cubic_distortions}) \cite{Vogt1999, Locherer1999_2, Howard2001}. The irreducible representation of the $M_1$ mode is two-dimensional, and so the order parameter of the transition, which we denote  as $Q$, has two components, $Q(q_1,q_2)$. The Landau free energy density $F_Q$ of a domain wall described by $Q$, then depends on both $q_1$ and $q_2$, as well as on the strain, $e$. It is given by the following expansion around the $\alpha_2$ bulk free energy density
\begin{align}
    F_Q(q_{1},q_{2},e) =& F_{\alpha_2}^{0} + a(q_{1}^{2} + q_{2}^{2}) + b(q_{1}^{2} + q_{2}^{2})^{2} \nonumber\\ 
    &+c(q_{1}^{2} + q_{2}^{2})^{3} 
    + dq_{1}^{2}q_{2}^{2} + e(q_1^4q_2^2+q_1^2q_2^4)  \nonumber\\
    &+ \lambda_{1}(q_{1}^{2} + q_{2}^{2})e_{s} + \lambda_{2}(q_{1}^{2} - q_{2}^{2})e_{as} \nonumber\\
    &+ \lambda_{3}(q_{1}^{2} + q_{2}^{2})e_{3}\nonumber\\
    &+ s [(\nabla q_{1})^{2} + (\nabla q_{2})^{2}] + C_{ij} e_{i} e_{j} \quad,
\label{eq:Landau_Ginzburg}
\end{align}
where we use Voigt and Einstein notations \cite{Hatch2003, Isotropy}. $F_Q$, $q_1$, $q_2$ and the strains are treated as continuous fields and the spatial coordinates $z$ (for example $q_1(z)$, etc.) are implied. The non-symmetry breaking ($e_s$) and symmetry-breaking ($e_{as}$) strains can be related to the amplitudes of the in-plane eigenvectors of the tetragonal elastic tensor, $e_1$ and $e_2$, through $e_s=e_1 + e_2$ and $e_{as}=e_1 - e_2$, and $e_3$ is the strain along the tetragonal axis. Parameters $a$, $b$ and $c$ describe the Landau potential up to sixth order in $q_1$ and $q_2$. Parameters $d$ and $e$ describe the additional coupling between order parameter components $q_1$ and $q_2$ up to sixth order that is not contained in $b$ and $c$. The parameters $\lambda_1$, $\lambda_2$ and $\lambda_3$ describe the separate couplings between the order parameter and the tetragonal strains. The Ginzburg parameter $s$ accounts for the variation of the order parameter $Q$ in the domain wall. The last term is the strain energy with $C_{ij}$ the elastic tensor. \\
To circumvent the explicit calculation of the strain dependence, we incorporate the energy-minimizing strains in effective Landau parameters for $q_1$ and $q_2$. The general free energy density for a domain wall described by $Q$ then simplifies to

\begin{align}
    F_Q = & F_{\alpha_2}^0 + 
    a_Q(q_1^2 + q_2^2) + b_Q(q_1^2 + q_2^2)^2 \nonumber\\
    &+ c_Q(q_1^2 + q_2^2)^3 + a_{2Q}q_1^2q_2^2 + b_{2Q}(q_1^4q_2^2+q_1^2q_2^4)\nonumber \\
    &+ s [(\nabla q_1)^2 + (\nabla q_2)^2].
\label{eq:Landau_Ginzburg_eff}
\end{align}
Parameters $a_Q$, $b_Q$ and $c_Q$ correspond to terms that are non-vanishing even when $Q$ has only one component (i.e. $q_2=0$) and parameters $a_{2Q}$ and $b_{2Q}$ describe the bidirectional coupling between the order parameter components that only occur when both $q_1$ and $q_2$ are non-zero.

\subsection{Extension of the Landau-Ginzburg free energy density to include the effect of additional charge}

Next we extend the Landau-Ginzburg free energy density to study the effect of additional charge, $\rho(z)$, introduced by reduction or doping.\\

\paragraph{Effect of charge on the order parameter, $Q$.} 

We begin by analyzing the effect of additional charge density on the $Q$ order parameter describing the $\alpha_2$ to $\beta$ transition by extending the free energy density expression as follows.

\begin{align}
   \nonumber F_Q =& F_{\alpha_2}^0 + a_Q(\rho)(q_1^2 + q_2^2)
    + b_Q(\rho)(q_1^2 + q_2^2)^2\\ \nonumber
    &+c_Q(\rho)(q_1^2 + q_2^2)^3 + a_{2Q}(\rho)q_1^2q_2^2  \\\nonumber
    &+b_{2Q}(\rho)(q_1^4q_2^2+q_1^2q_2^4)+ s(\rho) [(\nabla q_1)^2
    + (\nabla q_2)^2] \\
    &+\mu(\rho). 
\label{eq:Landau_Ginzburg_chem}
\end{align}

Here, the direct effect of the charge on the free energy density in the $\alpha_2$ reference structure appears explicitly as the chemical potential term $\mu(\rho)$. All additional effects of the change in chemical potential are incorporated in the $\rho$ dependence of the Landau parameters, $a_Q(\rho)$ etc. Note that we also account for the effect of charge on the gradient parameter $s(\rho)$.\\

\paragraph{Effect of charge doping on the amplitude of the cubic $R_5^-$ mode.}

In addition to affecting the $Q$ order parameter responsible for the $\alpha_2$ to $\beta$ transition, the addition of charge has the effect of reducing the amplitude of the cubic $R_5^-$ mode which is present in both the $\alpha_2$ and $\beta$ phases \cite{Walkingshaw2004, Wang2017}. This mode is the order parameter for the transition between the $\alpha_1$ and $\alpha_2$ structures (see Fig.~\ref{fig:phases_modes}). Complete suppression of the cubic $R_5^-$ mode therefore transforms the $\alpha_2$ phase to the higher symmetry $\alpha_1$ phase. In order to take this into account, we extend the Landau potential for $\alpha_2$ further by expanding this mode around its value in the $\alpha_2$ phase, $R_{\alpha_2}$. For convenience, we define an expansion parameter $R = R_{\alpha_2} - |R_5^-|$, which is zero in the $\alpha_2$ phase and increases with doping, reaching the value $R_{\alpha_2}$ in the $\alpha_1$ phase, and so its sign matches that of a conventional Landau theory order parameter. ($|R_5^-|$ is the  amplitude of the cubic $R_5^-$ mode at the particular doping value of interest). The new parameter $R$ therefore describes the {\it reduction} in the amplitude of $R_5^-$ in the transition from $\alpha_2$ to $\alpha_1$.

Including this degree of freedom in the Landau potential with this definition of $R$ leads to the free energy density: 

\begin{align}
    F_{QR} = & F_Q + a_R(\rho)R +  b_R(\rho)R^2 
    + c_R(\rho)R^3 + d_R(\rho)R^4 \nonumber\\
    &+ e_R(\rho)R^5 + f_R(\rho)R^6 
    + a_{RQ}(\rho)R(q_1^2+q_2^2) \nonumber\\
    &+ b_{RQ}(\rho)R^2(q_1^2+q_2^2) 
    + c_{RQ}(\rho)R^3(q_1^2+q_2^2) \nonumber\\
    &+ d_{RQ}(\rho)R(q_1^2+q_2^2)^2 
    + e_{RQ}(\rho)R^4(q_1^2+q_2^2) \nonumber\\
    &+ f_{RQ}(\rho)R^2(q_1^2+q_2^2)^2 
    + a_{R2Q}(\rho)Rq_1^2q_2^2 \nonumber\\
    &+ b_{R2Q}(\rho)R^2q_1^2q_2^2 
    + c_{R2Q}(\rho)R^3q_1^2q_2^2 \nonumber\\
    &+ d_{R2Q}(\rho)R(q_1^4q_2^2+q_1^2q_2^4) 
    + e_{R2Q}(\rho)R^4q_1^2q_2^2 \nonumber\\
    &+ f_{R2Q}(\rho)R^2(q_1^4q_2^2+q_1^2q_2^4) 
    + t(\rho) (\nabla R)^2.
\label{eq:Landau_Ginzburg_R}
\end{align}

Thus the additional energy, $F_{R}=F_{QR}-F_{Q}$, is 0 in the $\alpha_2$ phase where $R$ is 0. Note that we included $R2Q$ terms, in which $R$ is present in a single and $Q$ in two directions, up to eighth order.

\section{Computational details}

\subsection{Choice of exchange-correlation functional}

The properties of \ce{WO3} are unusually sensitive to the choice of exchange-correlation functional, with many studies in the literature suggesting different choices. Consistently good matches of relaxed structures to experimental structures have been reported using the B1-WC hybrid functional by Hamdi \textit{et al.} \cite{Hamdi2016, Garcia2012}, as well as by Wang \textit{et al.} using HSE-06 albeit not to the same degree  \cite{Wang2017}, but use of a hybrid functional is prohibitively expensive for our calculations. We found that the generalized gradient approximation (GGA) in the PBEsol implementation grossly underestimates the amplitude of the tetragonal $M_1$ mode; a similar underestimation of the oxygen rotations in GGA(PBE)-relaxed monoclinic \ce{WO3} has also been reported \cite{Kruger2012}.
A more detailed comparison with published  calculations is often problematic, since in many cases only the lattice parameters of relaxed \ce{WO3} bulk structures are reported but not the internal coordinates \cite{Wijs1999, Walkingshaw2004, Huda2008, Valdes2009, Wang2011, LambertMauriat2012, Ping2013, Hung2014, Ping2014, Saadi2014, AlvarezQuiceno2015, Gerosa2015, Mehmood2016, Pellegrini2019}. 

In this work we use the local-density approximation (LDA) description of the exchange-correlation functional. Our motivation is its good description of the amplitude of the tetragonal $M_1$ mode, which is the order parameter $Q$ of the $\alpha_2-\beta$ transition that we study in detail here. The lattice constants and phonon mode amplitudes of \ce{WO3} bulk structures that we calculate within the LDA in this work are listed in TABLES \ref{tab:lattice_parameters} and \ref{tab:phonon_modes}.

We note, however, that our chosen LDA implementation is not suitable for describing the  $\gamma$ domain walls because it does not yield a pronounced and necessary decrease in energy from the $\beta$ to the $\gamma$ phase as for instance reported by Hamdi \textit{et al}\cite{Hamdi2016}. A detailed discussion of this point is provided in the appendix section \ref{subsec:gamma_domain_wall}.

\subsection{Calculation of Landau-Ginzburg parameters}
\label{LGParameters}
The calculations to obtain the parameters of the Landau-Ginzburg free energies in Eqns.\ (\ref{eq:Landau_Ginzburg_chem}) and (\ref{eq:Landau_Ginzburg_R}) were performed using  the {\sc Quantum Espresso} (version 6.2.1) plane-wave pseudopotential DFT implementation \cite{Giannozzi2009, Giannozzi2017}. The choice of DFT implementation was made to be consistent with the electron-phonon coupling calculations, which we describe later. This forced us to use norm-conserving pseudopotentials as these were the only available option for electron-phonon calculations when this work was started. The norm-conserving LDA pseudopotentials were generated with the {\sc ONCVPSP} program and the input parameters provided by the PseudoDojo pseudopotential repository \cite{Hamann2013, Hamann_WP, VanSetten2018, PseudoDojo_GH}. A high cutoff energy of 120 Ry was necessary to converge the parameters in the Landau potentials, due to the use of norm-conserving pseudopotentials and the small core of the available W pseudopotential. We used valence electron configurations of $4f^{14} 5s^2 5p^6 5d^4  6s^2$ for the W atoms and $2s^2 2p^4$ for the O atoms.  The $k$- and $q$-point grid sizes were set to $12\times12\times12$ and $4\times4\times4$ respectively in the cubic phase, and scaled down relatively for larger unit cells, ensuring that they were always commensurate with each other as required for the electron-phonon calculations.

The Landau parameters were determined by calculating the energies of structures with different amplitudes of the $Q$ ($q_1$ and/or $q_2$) and/or $R$ distortions frozen into the reference $\alpha_2$ structure in the range from \SIrange[range-units=single]{0}{1.2}{\angstrom} per unit cell. We calculated the energies of a total of 360 distinct $(R,q_1,q_2)$ points, with the size of the unit cells allowed to relax in each case to satisfy the condition of energy-minimizing strain.

Parameters to sixth order (eighth order for the $R2Q$) were calculated for all described terms, with higher-order terms, constrained to be small and positive, included in the fit in each case to prevent unphysical negative divergence. 
To calculate the change of the parameters on charge doping, we repeated the set of calculations for a total of four different amounts of additional electrons up to 0.25 electrons per f.u. The cell parameters were set to those obtained from relaxations that did not contain additional charge and they were not allowed to relax further. The changes in the parameters were then fitted up to quadratic order of the charge density $\rho$ (see TABLE \ref{tab:Landau_parameters}).

We fitted the chemical potential $\mu(\rho)$ of the $\alpha_2$ reference structure with a quadratic dependence on the charge density $\rho$ in the free energy density (see TABLE \ref{tab:Landau_parameters}).
To accurately extract a value for the quadratic term, which is small compared to the linear term, a total of 80 energies for charge densities between 0 and 0.25 electrons per f.u. were calculated.

Finally, the gradient parameters $s$ and $t$ were obtained with the procedure described in the appendix section \ref{subsec:ginzburg_determination}. The inter-atomic force constants of the $\alpha_2$ phase in real space were calculated by interpolating the dynamical matrices on a $q$-grid. Force constant matrices were then interpolated for $q$-points on the $q$-paths $(1/2,1/2,0) \rightarrow (1/2+\delta, 1/2+\delta, 0)$ for the $M_1$ and $(0,0,0) \rightarrow (\delta, \delta, 0)$ for the $\Gamma_2^+$ mode (the corresponding $\alpha_2-\beta$ domain walls correspond to the $\left(110\right)$ crystallographic plane). The path length $\delta$ was set to 0.04. The branches belonging to the transition modes $M_1$ and $\Gamma_2^+$ in the force constant matrix dispersion were determined by symmetry combined with visual analysis of the respective displacements $\eta$. The corresponding gradient parameters $s$ and $t$ were finally obtained by performing a quadratic fit to the determined force constant branch as shown in equation (\ref{eq:sq}). To describe the change of the gradient parameters with charge, calculations of $s(\rho)$ and $t(\rho)$ were performed in the $\alpha_2$ cell with three values of $\rho$, and then fitted to fourth order in $\rho$ with the third order term  omitted (see TABLE \ref{tab:Landau_parameters}).

\subsection{Supercell calculations of domain wall structures}

\begin{figure*}[ht]
\begin{subfigure}{\textwidth}
\begin{tikzpicture}
    \node at (0,0)
        {\includegraphics[width=\textwidth, trim={0 3.6cm 0 14.5cm},clip]{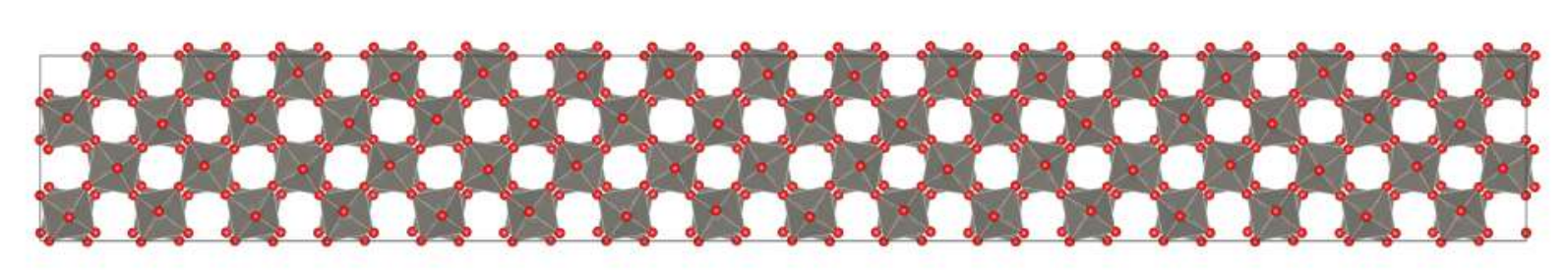}};
    \def \l {1.5}
    \draw[very thick, blue] (-6.62,-0.88) -- ++(45:\l) --++(135:\l) --++(225:\l) --cycle;
    \def \l {2.25}
    \draw[very thick, red] (-5.35,-1.26) -- ++(0:\l) --++(90:\l) --++(180:\l) --++(270:\l) -- cycle;
    \draw[very thick, red] (3.15,-1.26) -- ++(0:\l) --++(90:\l) --++(180:\l) --++(270:\l) -- cycle;
    \draw[very thick, black, dashed] (0,-1.5) -- (0,1.5);
    \draw[very thick, black, dashed] (-8.5,-1.5) -- (-8.5,1.5);
    \draw[very thick, black, dashed] (8.5,-1.5) -- (8.5,1.5);
    \draw[very thick, black] (4.25,-1.5) -- (4.25,1.5);
    \draw[very thick, black] (-4.25,-1.5) -- (-4.25,1.5);

    \draw[very thick, black!30!green, ->] (-6,-2) node[below] {$M_2^+$, $X_5^+$} -- ++(45:5.2);
    \draw[very thick, black!30!green, ->] (6,-2) node[below] {$M_2^+$, $X_5^+$} -- ++(135:5.2);
    \draw (0,-2)  node[below=2] {$M_2^-, R_5^-$} circle (0.16);
    \fill (0,-2)  circle (.04);
    \draw[very thick, black!30!green, ->] (-2.3,-2) node[below] {$M_2^-$} -- ++(135:5.2);
    \draw[very thick, black!30!green, ->] (2.3,-2) node[below] {$M_2^-$} -- ++(45:5.2);
    
    \draw[very thick, black, ->] (-1.5, 2) -- ++(3, 0) node[midway, above] {\large$\vv{z}$};
\end{tikzpicture}
\end{subfigure}
\caption{The initial unrelaxed supercell for the $\beta$ domain wall along a crystallographic $\left(110\right)$ plane (shown with the black dashed lines). We label the direction perpendicular to the plane of the wall as $\vec{z}$. The orthorhombic $\beta$ unit cell is indicated by the blue box and the ions that were kept fixed during the relaxation by red boxes (with black lines depicting their center). The cubic phonon modes and directions that are already present in the $\alpha_2$ phase are denoted by black vectors and those introduced by the $\alpha_2-\beta$ transition by green vectors.}
\label{fig:beta_wall}
\end{figure*}
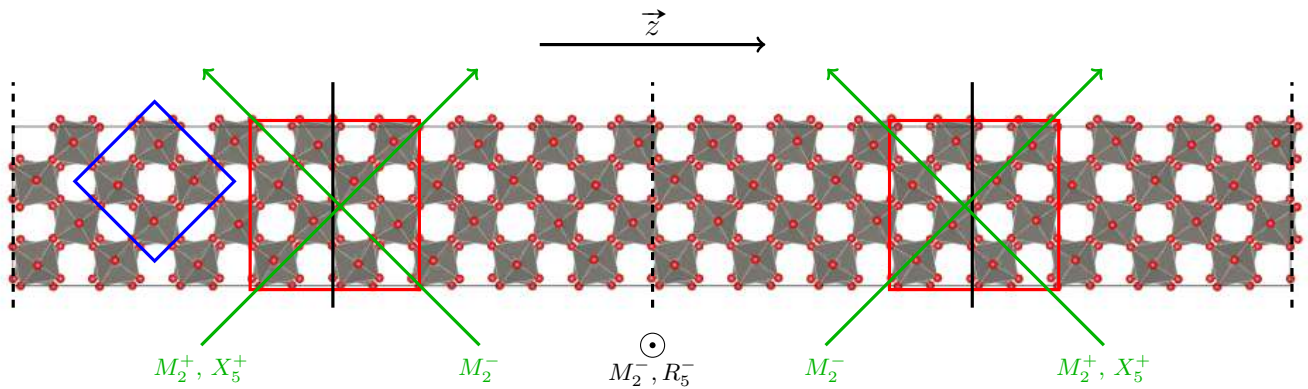

In addition to our Landau-Ginzburg calculations of the $\beta$ domain wall, we also performed direct calculations by explicitly relaxing the domain wall structure using DFT. We constructed a supercell containing two $\beta$ domain walls corresponding to $\left(110\right)$ crystallographic planes for subsequent relaxation (see Fig.~\ref{fig:beta_wall}). The supercell was generated as follows: First, the bulk structures were relaxed and one domain was constructed with the resulting relaxed structure. The second domain was then created by application of the point-group symmetry operations on the first domain that are lost during the transition. The supercell for the $\beta$ domain wall calculation contained 512 atoms and had dimensions of approximately $84\times11\times8$\SI{}{\angstrom}.

We then relaxed the supercell structure with some atoms fixed to the bulk structure (see Fig. \ref{fig:beta_wall}) as releasing the bulk cells causes a relaxation back to a single domain state. The relaxed structures were analyzed in terms of distortion modes. Additional charge was then introduced to the relaxed cells to determine if there was an accumulation of charge at the domain walls.

All calculations for the large domain wall supercells were performed with the GPU-accelerated {\sc VASP} (version 5.4.4) DFT implementation \cite{Kresse1996, Kresse1996_2, Hacene2012, Hutchinson2012}. The wave functions were expanded with a basis set of plane waves and their cutoff energy was set to \SI{800}{\electronvolt} for the bulk structures but was decreased to \SI{600}{\electronvolt} for the supercell relaxations. The core electrons were treated with projector-augmented waves, allowing for a lower plane-wave cutoff energy compared to the norm-conserving pseudopotentials employed for the rest of this work \cite{Kresse1994, Kresse1999}. The valence electron configurations were $5p^6 5d^4 6s^2$ and $2s^2 2p^4$ for the W and O atoms, respectively. The $k$-point grid sizes were chosen relative to a  $10\times10\times10$ grid for the cubic phase. The relaxation convergence criterion was \SI{E-3}{\electronvolt \per \angstrom} in the residual forces. Phonon mode amplitudes in the supercells were determined with the {\sc ISODISTORT} program from the {\sc ISOTROPY} Software suite \cite{Campbell2006, Isotropy}. \newline

\subsection{Calculations of electron-phonon coupling and superconducting critical temperature}

First-principles calculations of electron-phonon coupling were performed for electron-doped 8-atom $\alpha_1$ and 16-atom $\alpha_2$ cells. 
Electron doping was achieved by adding electrons with a compensating background charge rather than explicit inclusion of point defects; this method has been shown to describe well the charge-induced structural distortions in \ce{WO3} \cite{Walkingshaw2004}. For reasons discussed in the next subsection \ref{subsec:charge_relax}, we only relaxed the internal coordinates and the cell parameters were manually set by linear interpolation between the calculated values for the tetragonal \ce{WO3} and the cubic \ce{NaWO3} cells.

We used the {\sc EPW} package in conjunction with {\sc Quantum Espresso} \cite{Giustino2007, Ponce2016}. The cutoff energy had to be kept at an extremely large value of 120 Ry to converge the calculations, as tested by the convergence behaviour of the total electron-phonon coupling strength $\lambda$ in the cubic phase. Coarse $k$- and $q$-point grids were set relative to $12\times12\times12$ and $4\times4\times4$ grids of the cubic phase. Additionally, fine grid sizes for the $k$- and $q$-point grids were set relative to 200,000 and 100,000 random points for the cubic phase, respectively. W $5d_{xy}$, $5d_{xz}$ and $5d_{yz}$ orbitals were chosen for the wannierization procedure, which was performed using the {\sc wannier90} package \cite{Mostofi2014}. 

Superconducting critical temperatures were then extracted using the usual Eliashberg formalism of Bardeen-Cooper-Schriefer (BCS) theory, as described in Appendix section \ref{subsec:superconductivity}. The temperature for the Fermi occupations in Eqn. (\ref{eq:phonon_linewidth}) was set to \SI{0.075}{\kelvin} and the Fermi surface energy window of considered electron states was set to \SI{3}{\electronvolt}. The Coulomb pseudopotential parameter $\mu$ was set to 0.10. All results are given for smearings of \SI{0.05}{\milli\electronvolt} for the delta functions in Eqn. (\ref{eq:coupling_strength}) and \SI{0.25}{\electronvolt} for the frequency delta functions in Eqn. (\ref{eq:eliashberg}), respectively. None of the calculations were performed with the double-delta approximation.

\subsection{Lattice relaxation of charged unit cells}\label{subsec:charge_relax}
As the calculated stresses in charged unit cells with a constant background charge are not  well-defined in DFT implementations, we did not relax the volumes of our unit cells and supercells in the calculations for which we include additional electronic charge \cite{Bruneval2015}. 

To determine the validity of keeping the lattice parameters fixed in our Landau model, we compared our Landau parameters calculated using the relaxed lattice parameters of the undoped cells with calculations in which the cell parameters of the $\alpha_2$ cells were interpolated to those of the cubic \ce{NaWO3} cell. We found that the changes in the Landau curves caused by the volume change were negligible compared to those caused by the introduction of charge into the uncharged cells. Thus, the change in volume caused by doping within the doping range used here can be safely disregarded.

In the case of the electron-phonon calculations for the $\alpha_1$ phase, the change in cell size on doping could not be disregarded, as the phonon dispersions depend strongly on the cell volume. For example rotational modes are artifically stabilized if the cell volume is not allowed to increase, whereas antipolar modes are artifically destabilized. Thus, we opted for the compromise of linearly interpolating the lattice parameters between those calculated for the  tetragonal \ce{WO3} and for the cubic \ce{NaWO3} unit cells. We found this approximation to be sufficient to describe the correct general trends of the modes upon doping in the $\alpha_1$ phase. In particular, the rotational $R_5^-$ mode became softer with decreasing charge $x$ so that the $\alpha_2-\alpha_1$ transition occurred at a doping value close to the experimentally observed transition value of $x=0.2$ in \ce{Na_xWO_{3-x}} \cite{Shanks1974}. 
Also, the amplitude of the antipolar cubic $M_2^-$ mode decreased with increasing $x$ consistent with the literature \cite{Walkingshaw2004}. 

\section{Results of domain walls calculations}

\begin{table}
    \centering
    \begin{tabularx}{0.48\textwidth}{*{7}{>{\centering\arraybackslash}X}}
        \hline 
        \hline
        \multicolumn{1}{l}{$P4/nmm$} &
        \multicolumn{3}{c}{Current work} &
        \multicolumn{3}{c}{Experimental \cite{Locherer1999}} \\
        \hline
        & $a$ & $b$ & $c$ & $a$ & $b$ & $c$ \\
        & 5.314 & 5.314 & 3.872 & 5.303 & 5.303 & 3.935  \\
        & 5.282 & 5.282 & 3.872 & & &  \\

        \hline 
        \hline
        \multicolumn{1}{l}{$P4/ncc$} &
        \multicolumn{3}{c}{Current work} &
        \multicolumn{3}{c}{Experimental \cite{Vogt1999}} \\
        \hline
        & $a$ & $b$ & $c$ & $a$ & $b$ & $c$ \\
        & 5.272 & 5.272 & 7.833 & 5.276 & 5.276 & 7.846  \\
        & 5.178 & 5.178 & 7.750 & & &  \\
        
        \hline 
        \hline
        \multicolumn{1}{l}{$Pbcn$} &
        \multicolumn{3}{c}{Current work} &
        \multicolumn{3}{c}{Experimental \cite{Vogt1999}} \\
        \hline
        & $a$ & $b$ & $c$ & $a$ & $b$ & $c$ \\
        & 7.425 & 7.429 & 7.652 & 7.333 & 7.573 & 7.74 \\
        & 7.370 & 7.397 & 7.628 & & & \\
        
        \hline 
        \hline
        \multicolumn{1}{l}{$P2_1/n$} &
        \multicolumn{3}{c}{Current work} &
        \multicolumn{3}{c}{Experimental \cite{Vogt1999}} \\
        \hline
        & $a$ & $b$ & $c$ & $a$ & $b$ & $c$ \\
        & 7.438 & 7.404 & 7.613  & 7.303 & 7.538 & 7.692  \\
        & 7.356 & 7.401 & 7.622 & & & \\
        & $\alpha$ & $\beta$ & $\gamma$ & $\alpha$ & $\beta$ & $\gamma$ \\
        & 90 & 90.829 & 90 & 90 & 90.855 & 90 \\
        & 90 & 90.295 & 90 & & & 
    \end{tabularx}
    \caption{Lattice parameters (in \SI{}{\angstrom} and degrees) of relaxed structures obtained in this work with LDA-{\sc VASP} (upper rows) and LDA-{\sc Quantum Espresso} (lower rows) compared with experimental values from the literature.}
    \label{tab:lattice_parameters}
\end{table} 

\begin{table}
    \centering
    \begin{tabularx}{0.48\textwidth}{*{9}{>{\centering\arraybackslash}X}}
        \hline 
        \hline
        \multicolumn{1}{l}{$P4/nmm$} &
        \multicolumn{4}{c}{Current work} &
        \multicolumn{4}{c}{Experimental \cite{Locherer1999}} \\
        \hline
        & $X_5^+$ & $M_2^+$ & $M_2^-$ & $R_5^-$ & $X_5^+$ & $M_2^+$ & $M_2^-$ & $R_5^-$ \\
        & - & - & 0.239 & - & - & - & 0.253 & -  \\
        & - & - & 0.223 & - &   \\

        \hline 
        \hline
        \multicolumn{1}{l}{$P4/ncc$} &
        \multicolumn{4}{c}{Current work} &
        \multicolumn{4}{c}{Experimental \cite{Vogt1999}} \\
        \hline
        & $X_5^+$ & $M_2^+$ & $M_2^-$ & $R_5^-$ & $X_5^+$ & $M_2^+$ & $M_2^-$ & $R_5^-$ \\
        & - & - & 0.235 & 0.551 & - & - & 0.265 & 0.310  \\
        & - & - & 0.250 & 0.542 &   \\
      
        \hline 
        \hline
        \multicolumn{1}{l}{$Pbcn$} &
        \multicolumn{4}{c}{Current work} &
        \multicolumn{4}{c}{Experimental \cite{Vogt1999}} \\
        \hline
        & $X_5^+$ & $M_2^+$ & $M_2^-$ & $R_5^-$ & $X_5^+$ & $M_2^+$ & $M_2^-$ & $R_5^-$ \\
        & 0.099 & 0.224 & 0.239 & 0.474 & 0.284 & 0.322 & 0.253 & 0.353  \\
        & 0.112 & 0.281 & 0.253 & 0.462 &  &  &  &   \\
        
        \hline 
        \hline
        \multicolumn{1}{l}{$P2_1/n$} &
        \multicolumn{4}{c}{Current work} &
        \multicolumn{4}{c}{Experimental \cite{Vogt1999}} \\
        \hline
        & $X_5^+$ & $M_2^+$ & $M_2^-$ & $R_5^-$ & $X_5^+$ & $M_2^+$ & $M_2^-$ & $R_5^-$ \\
        & 0.085 & 0.251 & 0.243 & 0.514 & 0.262 & 0.339 & 0.246 & 0.400  \\
        & 0.125 & 0.303 & 0.251 & 0.466 & & & &
        
    \end{tabularx}
    \caption{ Total amplitudes of the main cubic phonon modes (in \SI{}{\angstrom}) in relaxed structures obtained with LDA-{\sc VASP} (upper rows) and LDA-{\sc Quantum Espresso} (lower rows) compared with experimental values from the literature. Phonon mode amplitudes were obtained with {\sc {\sc ISODISTORT}} \cite{Campbell2006, Isotropy}. The amplitudes correspond to the summed atomic displacements normalized to the cubic cell relaxed with LDA-{\sc VASP}.}
    \label{tab:phonon_modes}
\end{table}{}



\subsection{Landau-Ginzburg domain wall profiles}

Using the Landau-Ginzburg parameters obtained as described in Section~\ref{LGParameters}, we calculated the profiles of the order parameters across the domain walls by numerically minimizing the total free energy density functional given in equation (\ref{eq:Landau_Ginzburg_R}). The spatial grid of the order parameter fields, $z$, consisted of 251 points, spaced by $\Delta z=$  \SI{0.6}{\angstrom}. Self-consistent solutions were found as follows: For given $q_1(z)$ and $q_2(z)$ profiles at a specific total charge, the minimum energy charge distribution $\rho(z)$ was calculated. For this $\rho(z)$, the minimum energy $R(z)$ and subsequently $q_1(z)$ and $q_2(z)$ were obtained, after which the cycle was repeated. Self-consistency was achieved when the change in the total $z$-integrated energy density between steps was less than \SI{E-4}{\milli\electronvolt\per\angstrom\squared}. 

We checked our numerical approach for the simplified potential of eq. (\ref{eq:Landau_Ginzburg_simple}) by comparing with the analytical solutions of equations (\ref{eq:phi_analytical}) and (\ref{eq:Q_anlaytical}), and found excellent agreement between the numerical and analytical results. 

\begin{figure}
    \centering
    \includegraphics[width=0.49\textwidth]{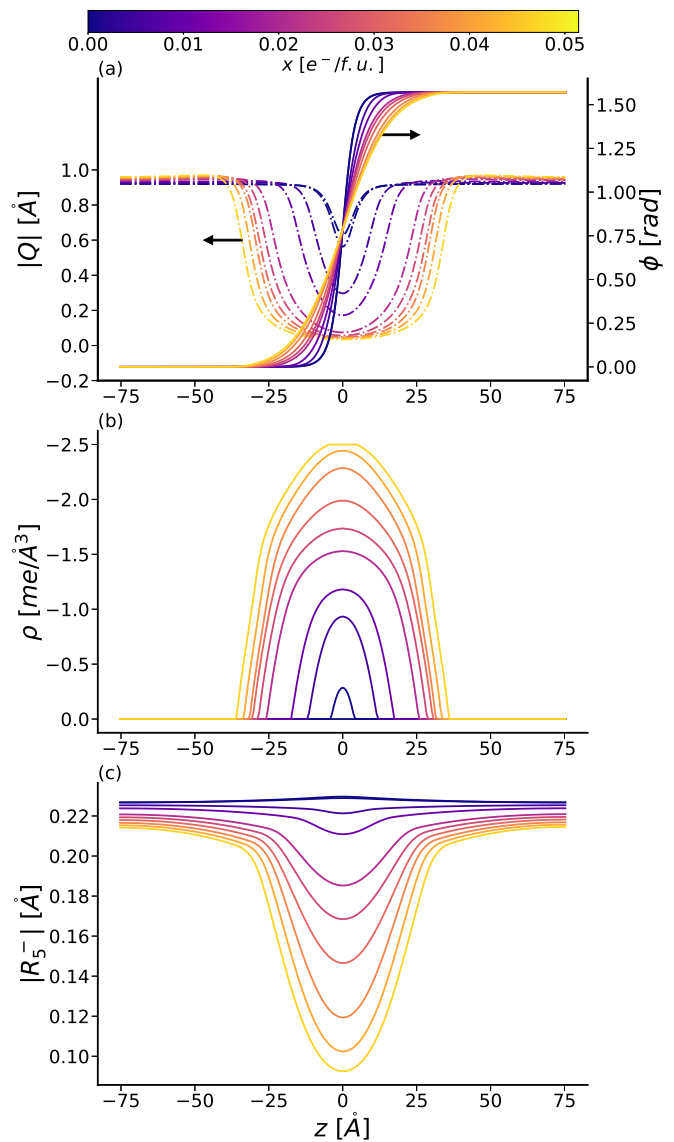}
    \caption{ Profiles of (a) the angle $\phi$ (solid lines) and amplitude $|Q|$ (dotted lines) of the mode $Q$, (b) the charge density $\rho$ (in milli-electronic charges per cubic angstrom) and (c) the total amplitude $|R_5^-|$ of the cubic $R_5^-$ mode in (c) across domain walls, calculated with the Landau-Ginzburg model. The total amount of additional charge corresponds to the doping level $x$ (in electrons per formula unit) depicted in the color bar.}
    \label{fig:Neel_wall}
\end{figure}

The calculated evolution of the order parameters across the energetically minimized $\beta$ domain walls for various doping levels is shown in Fig. \ref{fig:Neel_wall}. For clarity, we plot the magnitude $|Q|=\sqrt{q_1^2+q_2^2}$ and the angle $\phi=\arctan{q_1/q_2}$ of the order parameter.  Panel (a) shows the angle $\phi$ (solid lines) with respect to the $z$-axis along the wall. We see that, for the undoped case (blue line), the wall shows characteristic N\'eel-like behavior. The order parameter $Q$ retains 80 \% of its bulk amplitude across the wall and the reorientation is achieved by rotation of $Q$ along the wall as represented by $\phi$. The small reduction in amplitude at the wall can be explained by the bidirectional coupling of $Q$ (that is $q_1$ to $q_2$) which results in an energy reduction when the amplitude of $Q$ decreases. We extracted the wall width $2\xi$ by fitting  $Q(z)$ to a $|\tanh(z/\xi)|$ curve as in eq. (\ref{eq:phi_analytical}) and obtained a value of $\sim$\SI{1.35}{\nano\meter} for the undoped case. As can be seen from equations (\ref{eq:xi}) and (\ref{eq:xi2}), the widths are mostly determined by the gradient parameter $s$ of the order parameter $Q$ which is an order of magnitude larger than the Landau terms (see TAB. \ref{tab:Landau_parameters}). This value lies well within the general range of ferroelastic domain wall widths, which are generally between 0.2 and \SI{2}{\nano\meter} at low temperatures \cite{Salje2020}.

A distinct change in behavior is seen on introduction of electrons. At the most strongly doped example studied, 0.24 electrons per formula unit (yellow line), the wall is strongly Ising like, with the amplitude of $Q$ suppressed to zero in the wall region. At this highest doping level, the domain wall width is widened by a factor of around $3.6$ relative to the width in the undoped wall as measured by the fitting of the $\phi$ curves to $|\arctan(\exp(z/\xi))|$ curves. The crossover from undoped behavior to doped behavior, as well as the wall broadening, are gradual, with intermediate dopings (purple, red and orange colors) having intermediate behavior.

The origin of the evolution with doping is clear in Fig. \ref{fig:Neel_wall} panel (b) which shows the charge density as a function of position across the wall. We see that, for all doping levels, the charge accumulates in the wall region, and no additional charge remains in the bulk of the domains.

As discussed earlier, electron doping causes a reduction in the amplitude of the $Q$ order parameter, moving the structure towards the $\alpha_2$ phase. In addition, it causes a decrease of the $Z_3^+$ mode, parametrized by $|R_5^-|$ as shown in Fig.~\ref{fig:Neel_wall} panel (c). As a result the structure within the domain wall approaches that of the $\alpha_1$ phase. Note that we calculated Landau-Ginzburg parameters only for concentrations up to  \SI{-2.5}{\milli\elementarycharge\per\angstrom\cubed}. This is the origin of the forced cutoff of $\rho$ in the yellow curve in Fig.~\ref{fig:Neel_wall} panel (b). However, we expect that $|R_5^-|$ would decrease to 0 with increasing doping.


\subsection{Direct calculation of domain walls using density functional theory}

\begin{figure}
    \centering
    \includegraphics[width=0.5\textwidth]{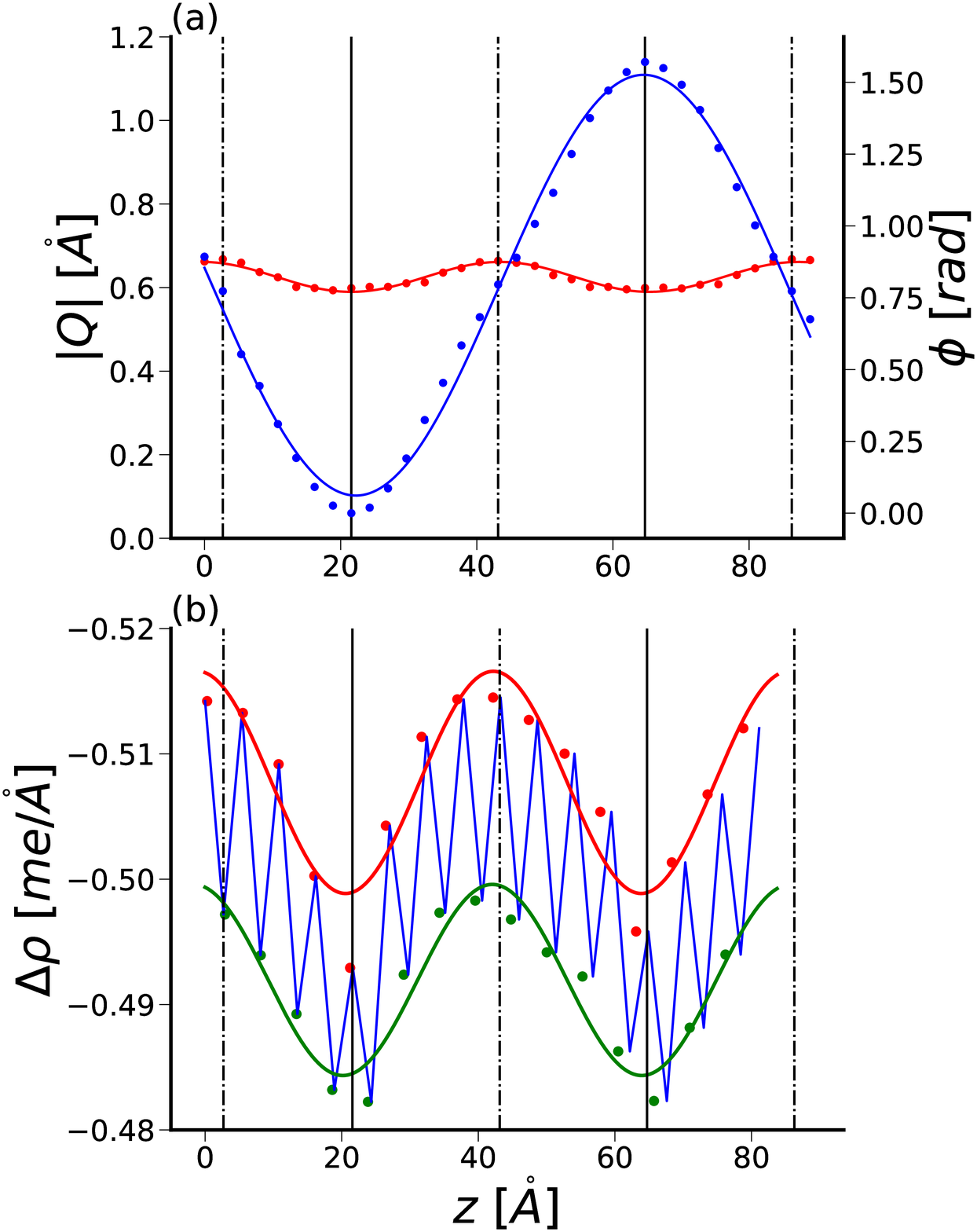}
    \caption{(a) Rotation angle $\phi$ (blue) and amplitude $|Q|$ (red) of the $Q$ mode along the [110] direction ($z$) of the relaxed $\beta$ domain wall supercell. (b) Change in macroscopic planar charge density $\Delta\rho$ (in milli-electronic charges per length(in \AA)) upon addition of charge in the relaxed $\beta$ domain wall structure. The red and green lines represent the upper and lower envelopes of the charge densities of red boxes as depicted in Fig. \ref{fig:beta_wall}. The dashed and solid lines in both panels indicate the positions of the fixed bulk unit cells and domain wall centers as in Fig.~\ref{fig:beta_wall}. }
    \label{fig:DW_BB}
\end{figure}

Motivated by our estimation of the domain wall widths of $\sim$14~\AA{} from our Landau-Ginzburg model, we next performed a full density functional calculation of the domain wall structure shown in Fig.~\ref{fig:beta_wall}. 

Our calculated layer-resolved order parameter angle $\phi$ and its amplitude $|Q|$ are shown in Fig. \ref{fig:DW_BB} panel (a) as a function of position perpendicular to the wall plane, $z$. Consistent with our results from Landau-Ginzburg theory, we find the domain wall to be predominantly of the N\'eel type as represented by the gradual transition in $\phi$. 
In Fig. \ref{fig:DW_BB} panel (b) we show the calculated charge density distribution obtained by adding an additional but small electronic charge of $1.6\times 10^{-2}e$ to the supercell. We find that, as in our Landau-Ginzburg simulations, the charge accumulates at the walls. While the macroscopic planar charge density shows an alternating behaviour from site to site, there is a clear depletion of charge from the bulk towards the domain wall structure as indicated by the top and bottom envelopes of the density. We note that a definitive study would require further relaxation of the wall after the introduction of the charge.
In addition, a systematic study of larger supercells would be desirable to ensure that there are no interactions between the walls and that full convergence to the bulk values is achieved in the intermediate regions. The supercells used here were barely large enough to host two domain walls as we observe no clear bulk plateau in the order parameters.

A clear difference compared to the Landau-Ginzburg model can be found in the amplitude of the $Q$ ($M_1$) mode in the domain walls. In the DFT-calculated walls we observe a slight increase in $Q$, whereas it decreased in the walls obtained from Landau-Ginzburg theory. The reason for this is the limitation of the phase space used in the Landau-Ginzburg model.
When comparing fully relaxed structures of the $Pbcn$ ($\beta$) (with order parameter direction $Q(a,0)$) and the $P\overline{4}2_1c$ (corresponding to order parameter direction $Q(a,a)$) phases we observe that the latter has a lower energy and a higher total $Q$ amplitude than the former.  Thus, we expect that including additional order parameters in the Landau-Ginzburg model would also lead to an increase in $Q$ in the center of the domain wall. However, this would lead to a highly increased dimension of the phase space, making the Landau-Ginzburg parameterization unfeasible. Due to the small amplitude of these additional distortions, it is reasonable to assume that the difference in domain wall width would be minor if they were included.\newline Furthermore, we expect the inaccuracy in the $Q$ displacement to be less relevant in the charged domain walls, as the charge reduces the amplitude of both $Q$ and $R_5^-$ at the domain wall.

\subsection{Summary of domain wall results}

In summary, we investigated the structure of the $\beta$-type domain walls that form during the phase transition from the tetragonal $\alpha_2$ phase to the orthorhombic $\beta$ phase in WO$_3$ in the framework of Landau-Ginzburg and density functional theories. Our Landau-Ginzburg calculations showed that the ferroelastic walls in the undoped case are mostly N\'eel-like, with the amplitude of the order parameter $Q$ retaining ${\sim}80\%$ of its bulk value. We found the domain wall width $2\xi$ to be around 14~\AA~in the undoped case. Electronic doping increased our calculated domain wall width and led to an accumulation of the additional charge in the domain wall. The domain wall width of \ce{WO3} $\beta$ domain walls at very low temperatures has been reported to be around $2w=$ \SI{1.2}{\nano\meter} in experiments \cite{Locherer1998, Salje2004}, which is very close to the value suggested by our Landau-Ginzburg model. We found that the charge accumulation at the walls caused an increasingly large Ising-type component, indicated by the drop in the order parameter amplitude $|Q|$ across the wall. The accumulated charge also reduced the magnitude of the $|R_5^-|$ order parameter, so that the structure approached that of the $\alpha_1$ phase in the wall region. Using DFT calculations on supercells, we were able to confirm the N\'eel-type character of the domain walls, as well as the predicted accumulation of charge at the domain walls.


\section{Implications for domain wall superconductivity}

Motivated by our finding that the charge accumulates at the domain walls and causes a local $\alpha_1$-like structure, we next study the superconducting properties of this phase. Experimentally, the $\alpha_1$ phase of \ce{Na_xWO_{3-$x$}}, which also has $P4/nmm$ symmetry \cite{TRIANTAFYLLOU1997479}, was shown to be superconducting. Similar to other superconducting tungsten bronzes, the superconductivity shows two general features. First, for each dopant type, superconductivity occurs in only one high-symmetry structure. For smaller alkali metals (Na and K) these are structures of tetragonal symmetry, while for larger alkali metals (Rb and Cs) the structures are hexagonal. At doping levels $x$ that lie above or below the $x$-range of these phases, superconductivity is not found. The second feature is a decrease in $T_c$ with increasing $x$, within the superconducting phase. Thus, the highest $T_c$ is reported at the lowest $x$-value at which the superconducting phase is still retained; lower doping results in a phase transition to the lower-symmetry, non-superconducting phase. Both properties implicate the soft mode associated with the corresponding structural phase transition in the superconductivity mechanism \cite{Sweedler1965, Shanks1974, Ngai1976, Aird1998}.

Our approach is to calculate and analyze $T_c$ as a function of doping within standard Bardeen-Cooper-Schrieffer (BCS) theory \cite{BCS1957} for the $\alpha_1$ phase of WO$_3$. While BCS theory has been shown to capture some aspects of the behavior of doped \ce{WO3} \cite{Shanks1974, Stanley1979,Aird1998_2,Reich2000}, the absence of superconductivity in the $\alpha_2$ phase, and the decrease in $T_c$ with increasing doping are not well understood (the latter has even been described as the 
``$T_c$ paradox''! \cite{Shanks1974, Ngai1976, Ngai1978, Brusetti2002, Brusetti2007, Bocarsly2013, Pellegrini2019}), and we explore these aspects here. 

We calculate the electron-phonon coupling matrix,
\begin{equation}
g_{nm}^{\nu} (\vec{k},\vec{q}) = \left\langle \varphi_{m\vec{k}+\vec{q}} \left| \Delta_{q}^{\nu} V_{KS} \right| \varphi_{n\vec{k}} \right\rangle ,
	\label{eq:ep_matrix}
\end{equation}
using density functional perturbation theory (DFPT) \cite{Baroni2001}. Here, $\Delta_{q}^{\nu} V_{KS}$ is the phonon perturbation to the Kohn-Sham potential, and the matrix elements are the 
transition probability amplitudes for an electron in initial state $\varphi_{n\vec{k}}$ with wave vector $\vec{k}$ and band $n$, scattering to final state $\varphi_{m\vec{k}+\vec{q}}$ of band $m$, via a phonon of wave vector $\vec{q}$ and branch $\nu$. We then evaluate the superconducting critical temperature using the semi-empirical Allen-Dynes equation\cite{Allen1975}
\begin{gather}
	T_c = \dfrac{\left\langle \omega\right\rangle }{1.2}\exp \left[ \dfrac{-1.04(1+\lambda)}{\lambda - \mu(1+0.62\lambda)}\right],
	\label{eq:Allen_Dynes}
\end{gather}
with the coupling strength, $\lambda$, and the weighted phonon frequency $\left<\omega\right>$, extracted from the electron-phonon matrix as described in the appendix section \ref{subsec:superconductivity} and an empirical value of 0.1 taken for the effective Coulomb repulsion $\mu$.

\subsection{BCS theory applied to the bulk $\alpha_1$ phase}

\begin{figure}
    \centering
    \includegraphics[width=0.5\textwidth]{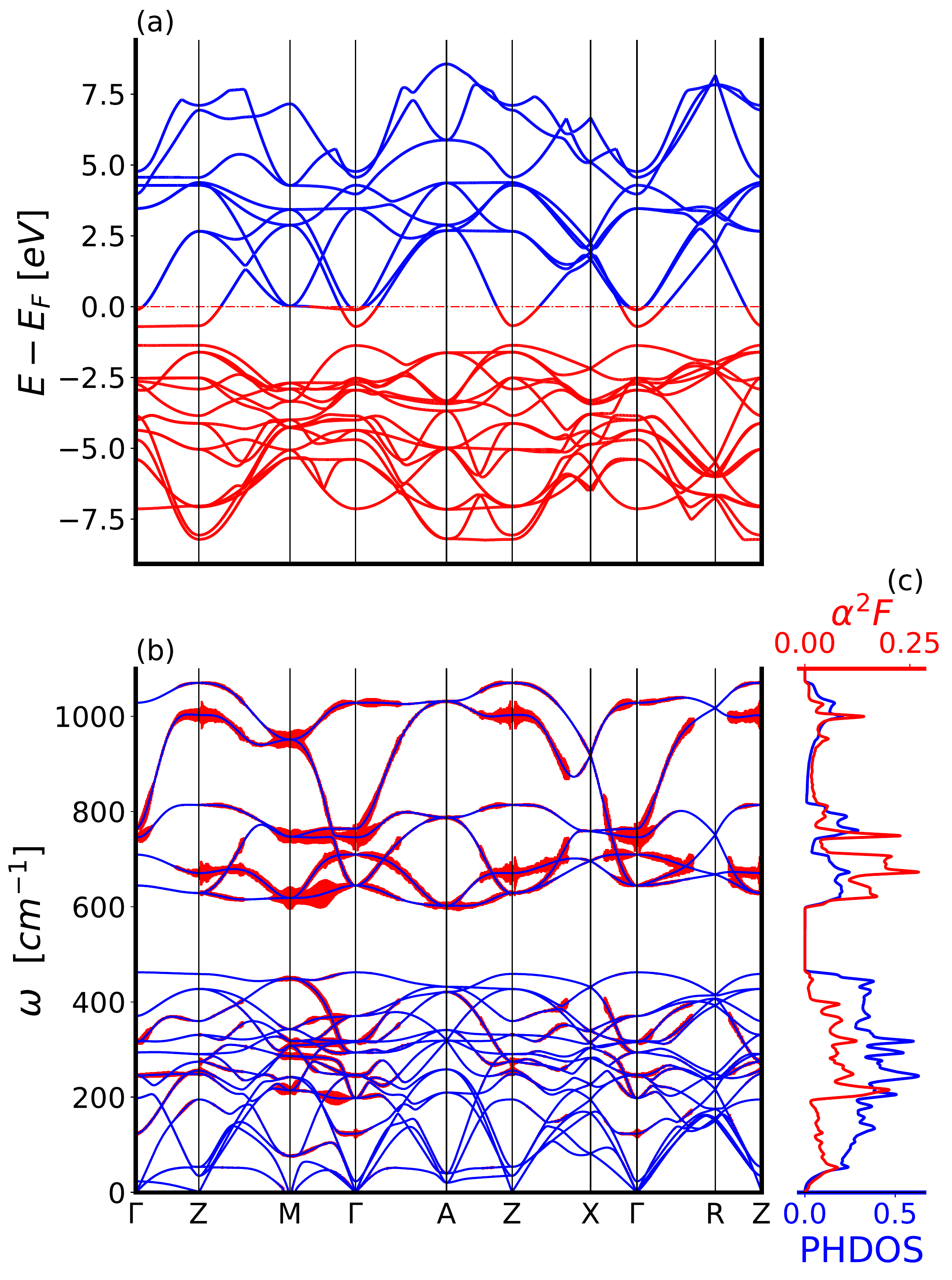}
    \caption{Calculated $\alpha_1$ electron bands (a), phonon bands and phonon linewidths (b), phonon density of states and $\alpha^2F$ (c) for the case of $x=0.125$ additional electrons per formula unit. The phonon linewidths are shown as vertical red bars in the phonon band structure plot and they are scaled by a factor of 10 for visibility. Occupied states in the electron band structure are shown in red, unoccupied states in blue, and the horizontal line at zero eV is the Fermi level.}
    \label{fig:phlw}
\end{figure}

\begin{figure}
    \centering
    \includegraphics[width=0.5\textwidth]{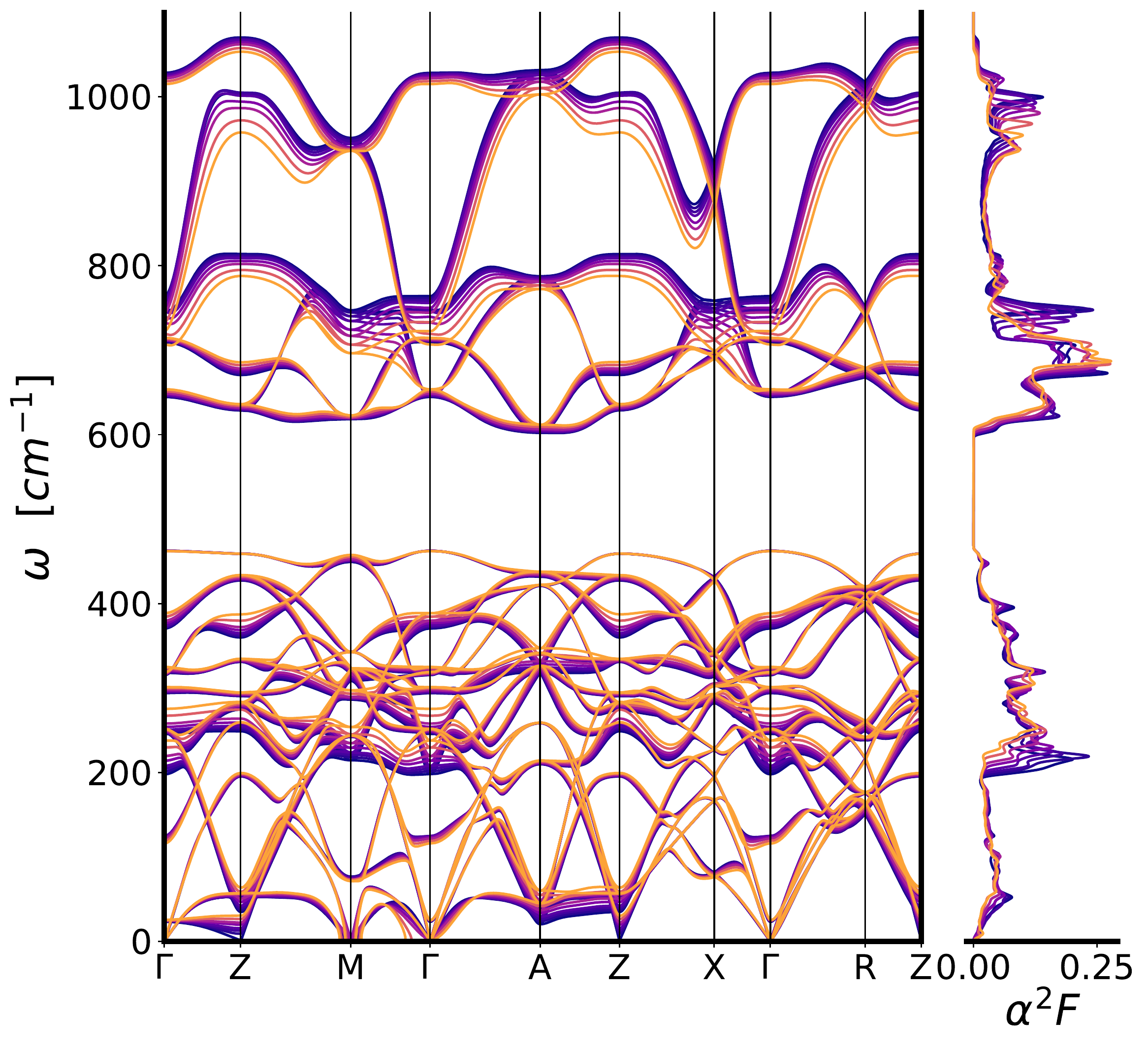}
    \caption{Calculated phonon bands (left) and $\alpha^2F$ (right) for the $\alpha_1$ structure, for a range of added electron concentrations. Blue corresponds to the lowest (0.125 electrons per formula unit) and orange to the highest (0.24 electrons per formula unit) doping levels, with successive lines corresponding to successive points in the plots of Fig.~\ref{fig:Tc}.}
    \label{fig:ph_e}
\end{figure}

\begin{figure}
    \centering
    \includegraphics[width=0.35\textwidth]{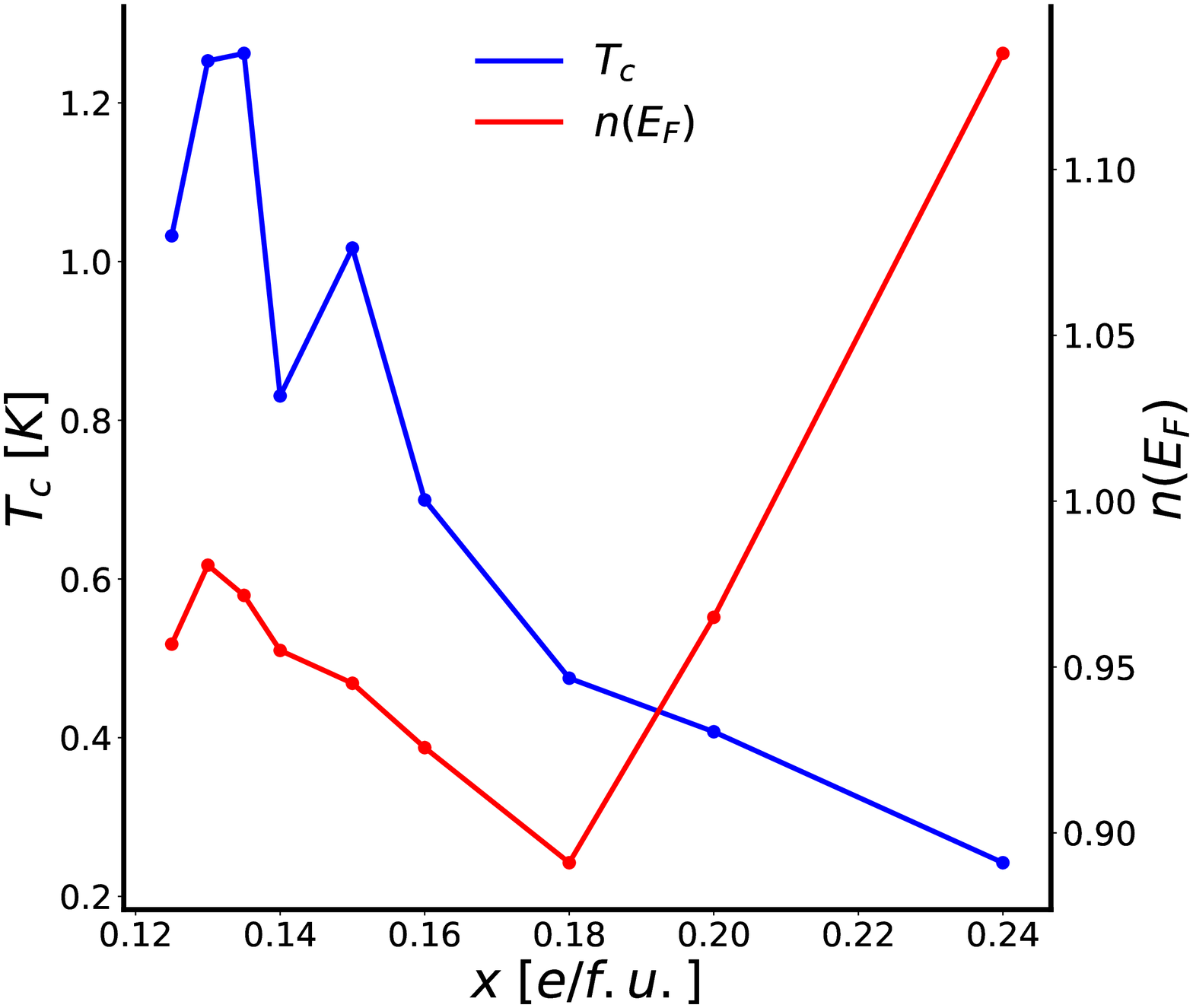}
    \caption{Calculated superconducting critical temperature $T_c$ and electron density of states at the Fermi level $n(E_F)$ in the $\alpha_1$ phase, as a function of added electrons per formula unit, $x$.}
    \label{fig:Tc}
\end{figure}

We begin by calculating the superconducting $T_c$ for the $\alpha_1$ phase, to see whether the measured decrease in $T_c$ with increasing doping is correctly captured within BCS theory. The $\alpha_1$ structure is stable for calculated electron concentrations larger than $x \approx 0.125$ (for lower concentrations, it has an unstable $Z_3^+$ mode, indicating the transition to the lower-energy $\alpha_2$ structure), which is therefore the lowest doping concentration that we consider. Our calculated electron bands, phonon bands, phonon linewidths, phonon density of states and Eliashberg spectral function $\alpha^2F$ for $x=0.125$ are shown in Fig.~\ref{fig:phlw}, where the soft $Z_3^+$ mode is the lowest frequency $Z$ mode in the phonon bands close to zero frequency. Interestingly, while there is some electron-phonon coupling at low frequency, it is considerably stronger at higher frequencies, with the highest values of $\alpha^2F$ occuring at around 600 - 800 cm$
^{-1}$. This suggests that, at least in the BCS picture, the soft mode is not the most relevant in determining the superconducting $T_c$. The subsequent changes in the phonon bands and $\alpha^2F$ upon increase of $x$ are presented in Fig.~\ref{fig:ph_e}. We see that, as expected, the $Z_3^+$ soft mode hardens with increasing doping, leading to a reduction of $\alpha^2F$ at low frequency with increasing doping. Interestingly, the high energy phonons shift to lower frequencies as doping is increased, with corresponding shifts of the peaks in $\alpha^2F$ to lower frequency. Finally, the calculated superconducting critical temperature, $T_c$, and density of states at the Fermi level, $n(E_F)$, are shown in Fig.~\ref{fig:Tc} as a function of doping concentration, $x$ \footnote{
Note that we excluded the imaginary frequencies around $M$ ($q=[1/2, 1/2, 0]$) in our calculation of $\alpha^2F$ in the integration in equation (\ref{eq:eliashberg}). We also checked the influence of the adjacent real values that are close to zero by excluding the real part of the branch within a window bounded by a maximum frequency of \SI{50}{\per\centi\meter} and a box around $M$ defined by $\Delta q=0.2$ in each direction. We found that including or excluding the phonons in this window causes only a small change in the calculated $\alpha^2F$ (see appendix Fig.~\ref{fig:a2F_change}) between the two cases.}. 
The first points in Fig.~\ref{fig:Tc} (at $x=0.125$) correspond to the band structures shown in Fig.~\ref{fig:phlw}. 

{}

{}

{}

Fig.~\ref{fig:Tc} clearly illustrates that our BCS-theory calculations reproduce the experimental trend of  decreasing $T_c$ with increasing doping, with the calculated  maximum in $T_c$ at $x=0.125$ coinciding with the calculated $\alpha_1-\alpha_2$ transition, where the transition $Z_3^+$ mode starts to become imaginary. In addition, our calculated $T_c$'s are comparable to the reported values ($\lesssim$ \SI{2}{\kelvin} in bulk samples), although we emphasize that their actual magnitudes should not be over-interpreted, since they are sensitive to the spreads in $\alpha^2F$ integration in equation (\ref{eq:eliashberg}) and the value of the screened Coulomb potential $\mu$ in the Allen-Dynes formula (\ref{eq:Allen_Dynes}). The trend of a decreasing $T_c$, however, is robust to these parameters. Therefore we conclude from our calculations that conventional BCS theory captures the observed evolution of $T_c$ with doping in the $\alpha_1$ phase of WO$_3$.\footnote{We note that a recent paper \cite{Pellegrini2019} using ostensibly similar methods obtained an increase in $T_c$ with increasing doping. We have been unable to reproduce the results of their work or understand the origin of the difference with our calculations. Note that they do reproduce the experimental behavior when they dope by introducing point defects rather than with electrostatic doping.} 

Given the good agreement of our computational BCS theory results with experiments, we next analyze them to rationalize the behavior. In particular, the decreasing $T_c$ with $x$ was unexpected within a simple BCS picture, since the electron density of states at the Fermi level, $n(E_F)$, has been reported from photoemission measurements to increase with increasing $x$ in tetragonal \ce{Na_xWO_{3-x}} \cite{Hochst1982, Egdell1982}. Since the BCS Cooper-pair binding energy scales as $\exp\left[ -1/(n(E_F) V_{ep}) \right]$ ($V_{ep}$ is the inter-electronic attraction caused by the electron-phonon coupling),  an increase in $n(E_F)$ should in turn lead to an increase in $T_c$, provided that the electron-phonon coupling strength remains constant with electron density. Our calculations of $n(E_F)$ (Fig.~\ref{fig:Tc}) in fact indicate that, within the $\alpha_1$ phase, $n(E_F)$ (red line) at first decreases with increasing doping (as does $T_c$). Note that there is no inconsistency with Refs.~\onlinecite{Hochst1982} and ~\onlinecite{Egdell1982}, which provided measured $n(E_F)$ values only above $x=0.25$.  At higher dopings (above $x=0.18$) $n(E_F)$ starts to increase, while $T_c$ continues to decrease. This lack of correlation between $T_c$ and $n(E_F)$ points to a doping dependence of the electron-phonon matrix elements.

\begin{figure}
    \centering
    \includegraphics[width=0.35\textwidth]{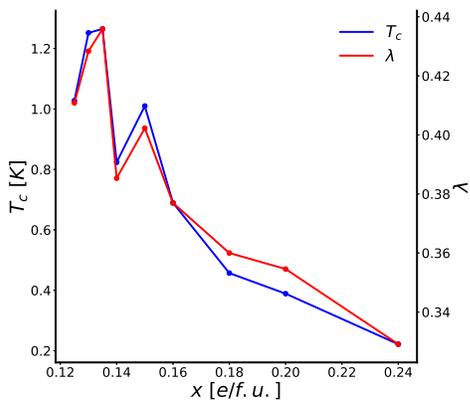}
    \caption{Calculated superconducting temperature $T_c$ and total coupling strength $\lambda$ as a function of added electrons per formula unit, $x$, in the $\alpha_1$ phase.}
    \label{fig:lambda_Tc}
\end{figure}

In Fig.~\ref{fig:lambda_Tc} we show our calculated doping dependence of $T_c$ and total coupling strength, $\lambda$, as defined in equation (\ref{eq:lambda}). First we note that, over the whole range, the coupling strength, $\lambda$, has a substantial value, consistent with the measurable superconductivity. Second, as we expected, it is clear that the value of $\lambda$ decreases with increasing doping, explaining the corresponding decrease in $T_c$ according to the Allen-Dynes formula given in equation \ref{eq:Allen_Dynes}. In particular, the calculated $T_c$ tracks closely the calculated value of $\lambda$.

Finally, to understand the change in superconductivity across the $\alpha_1-\alpha_2$ transition, we performed a calculation of $T_c$ in the $\alpha_2$ phase where experimentally superconductivity has not been measured. We chose a value of $x=0.125$ for the $T_c$ calculation in the $\alpha_2$ phase, and adjusted the lattice constants as outlined in section \ref{subsec:charge_relax} for the $\alpha_1$ phase, so that the resulting system in the $\alpha_2$ phase was quite far from the $\alpha_1-\alpha_2$ transition. (The amplitude of the transition mode $Z_3^+$ was \SI{0.51}{\angstrom}, compared with the amplitude of \SI{0.76}{\angstrom} in the undoped $\alpha_2$ phase). 
 As expected, our calculated $T_c$ dropped sharply from the calculated $\alpha_1$ value, to \SI{0.018}{\kelvin}, consistent with a sharp drop in the calculated $\lambda$ value to 0.25. The bands and linewidths for the $\alpha_2$ case are shown in appendix Fig.~\ref{fig:phlw2}.

\begin{figure}
    \centering
    \includegraphics[width=0.35\textwidth]{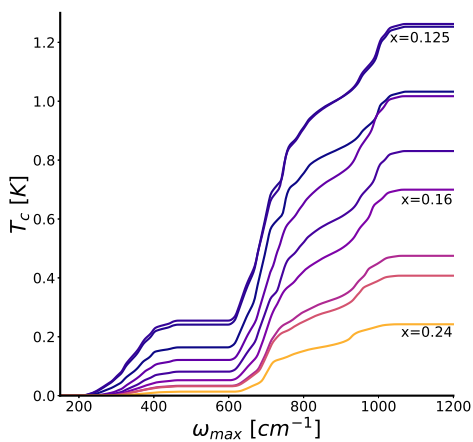}
    \caption{Calculated $T_c$ as a function of the upper integration frequency limit, $\omega_{max}$, in equations (\ref{eq:lambda}) and (\ref{eq:freq}). Blue curves show the lowest and orange curves the highest doping $x$ (in electrons per formula unit), over the same range as in Fig. \ref{fig:Tc}.}
    \label{fig:Tc_w}
\end{figure}

\subsection{Discussion of bulk superconductivity results}

While our calculations indicate that the superconducting behavior of WO$_3$ can be reproduced within standard BCS theory, this is of course not definitive evidence that WO$_3$ is a BCS superconductor. In this section we discuss two other models for superconductivity -- based on soft modes and bipolarons respectively --  that have been discussed in the literature.

We begin with a discussion of the importance of the soft mode, whose strong change in frequency with doping was originally proposed to account for the apparently paradoxical behaviour of $T_c$ upon doping \cite{Shanks1974}, in spite of its absence in inelastic neutron scattering experiments \cite{Ngai1976, Ngai1978}. As mentioned above, our calculated phonon linewidths and doping dependence of $\alpha^2F$ point to a small, if any, role of the soft mode at the BCS level; here we quantify its contribution. In Fig.~\ref{fig:Tc_w}, we plot the calculated $T_c$ as a function of the maximum frequency of the phonons included in the calculation, for a range of doping values within the $\alpha_1$ phase.  We see that the modes below \SI{200}{\per\centi\meter}, which include the soft mode, contribute negligibly to $T_c$. As noted above, the modes above around \SI{600}{\per\centi\meter}, which had the largest phonon bandwidth and the strongest frequency shifts on doping, contribute most to $T_c$ at every doping concentration.\footnote{We note that again our findings are in contrast to those of Ref.
~\onlinecite{Pellegrini2019}, who report that 60 \% of the electron-phonon coupling comes from the lowest quarter of the spectrum.}  While our calculations were performed for tetragonal structures, we note that Brusetti \textit{et al.} have attributed the increase in $T_c$ with decreasing $x$ in {\it hexagonal} \ce{Rb_xWO_{3-x}} to changes in  electron-phonon coupling for phonons with a frequency of more than \SI{240}{\per\centi\meter} \cite{Brusetti2007}.

Before leaving the topic of soft-mode superconductivity, we point out that the superconductivity in WO$_3$ is somewhat reminiscent of that in SrTiO$_3$, for which a model of superconductivity mediated by fluctuations associated with the ferroelectric quantum critical point has been proposed \cite{Edge2015}. The quantum criticality model had considerable success in reproducing the measured behavior, as well as in making rather bold predictions about strain and isotope effects that were subsequently verified experimentally \cite{Stucky2016, dunnett2018strain, Herrera2019, Schumann2020}. An important difference is that SrTiO$_3$ has a superconducting dome as a function of doping, whereas in WO$_3$ an analogous picture would have the left side of the dome cut off due to the absence of superconductivity in the $\alpha_2$ phase. If this mechanism is relevant in WO$_3$, a large and anomalous oxygen isotope effect on $T_c$ should be observed.  

Second, we note that electronic carriers in doped WO$_3$ have been shown, using optical absorption, conductivity and electron spin-resonance data, to form both polarons -- in some cases in combination with free carriers \cite{SALJE1979237,saljeguettler,Ruscher_1988} -- and bipolarons \cite{Schirmer1980, Schirmer1980_2}. While it was speculated that bipolarons could be responsible for the high temperature surface superconductivity in H$_x$WO$_3$ \cite{Reich1999}, 
their role in mediating superconductivity in tungsten-based oxides was subsequently largely neglected until a recent measurement of sheet superconductivity in the shear planes of the \ce{WO_{2.90}} Magneli phase \cite{Shengelaya2020}. The remarkably high reported $T_c$ of \SI{80}{\kelvin} in this system was attributed to  W$^{5+}$-W$^{5+}$ bipolarons, which were identified using electron paramagnetic resonance.
A recent density functional study of self-trapped polarons in \ce{WO3} succeeded in capturing a polaronic state, with substantial lattice distortions, in the simulations \cite{Bousquet2020}, although the polaron was at higher energy than the delocalized electron. The role of electron localization and its coupling to the lattice is clearly an important area for future study \cite{Salje2020}.

\subsection{From bulk to sheet superconductivity}

To link our bulk \ce{WO3} results to the sheet superconductivity reported at the \ce{WO3} $\beta$ domain walls \cite{Aird1998, Aird1998_2}, we revisit the $\beta$ domain walls that we obtained from Landau-Ginzburg theory and our density functional calculations. We can make three main inferences. First, from both studies we see that it is lower energy for electronic charge to be at the domain than in the surrounding $\beta$ bulk structure leading to local charge accumulation at the walls. Second, this local increase in charge induces a local transition to the tetragonal bulk $\alpha_1$ phase in the domain walls. And third, the additional charge, combined with the presence of the $\alpha_1$ phase, leads to strong enough electron-phonon coupling to enable superconductivity in the domain walls.

Many of the samples in which domain wall superconductivity was measured showed a stripe pattern of parallel ferroelastic domain walls of only one type (see for example Ref. \onlinecite{Aird1998_2}). Since similar samples were characterized in detail and shown to consist of $\beta$ domain walls \cite{Locherer1998}, it is likely that the superconducting samples contained only $\beta$ domain walls. Whether $\gamma$ domain walls are also superconducting, and if so by what mechanism, is an interesting open question for future study. 

\section{Conclusion}

In this work, we calculated the structure and properties of the ferroelastic domain walls within the $\beta$ phase of WO$_3$, using a combination of first-principles density functional calculations and Landau-Ginzburg theory. We showed that the ferroelastic $\beta$ domain walls have mixed N\'{e}el and Ising character, and found that free electronic charge preferentially accumulates at the domain walls. We showed that this accumulation of charge leads to a broadening of the walls and an increase in their Ising character, as well as a change in the atomic structure within the domain wall structure to the $\alpha_1 (P4/nmm)$ phase. This latter phase is known to be the superconducting phase in doped \ce{WO3}, suggesting that the domain wall superconductivity is a consequence of the combined electron accumulation and local structural change at the walls. 

To investigate further this possible link between domain wall and bulk superconductivity, we performed electron-phonon calculations based on DFT to calculate the $T_c$ as a function of doping in the bulk $\alpha_1$ phase at the BCS-theory level. Our calculated values were comparable in magnitude to the measured values ($\lesssim$\SI{2}{\kelvin}) and showed the same trend of decrease in $T_c$ with increasing doping. The evolution of $T_c$ with doping correlated with a reduction in the electron-phonon coupling, with the largest contribution coming from high frequency phonons above approximately \SI{600}{\per\centi\meter}.

Our calculations suggest that the superconductivity at the domain walls in WO$_3$ results from the combined accumulation of charge at the walls and the structural changes at the domain walls that are induced by the presence of the carriers.

\section*{Acknowledgements}

This work was supported by by the K\"orber Foundation and the ETH Z\"urich. Calculations were performed at the Swiss National Supercomputing Centre (CSCS) under project IDs s889 and eth3 and on the {\sc Euler} cluster of {\sc ETH Zurich}. A.N. acknowledges support from the start-up grant at the Indian Institute of Science (Grant number: SG/MHRD-19-0001). We acknowledge helpful discussions with Eric Bousquet who made us aware of the ill-defined pressures in charged unit cells in DFT implementations. Data and data analysis presented in this work can be found on \url{https://github.com/noemas/WO3}.

\section{appendix}

\subsection{Analytical solutions of domain wall profiles}\label{subsec:analytical_solutions}
Using the Landau-Ginzburg free energy density expression, one can calculate domain wall profiles by minimizing the free energy density with appropriate boundary conditions. For a 2D order parameter, in general, we can have two types of domain walls, N\'{e}el-type (order parameter rotates along the wall) and Ising-type (order parameter vanishes on the domain wall). These two limiting cases can be calculated analytically for a simple Landau theory of the form:
\begin{equation}
\begin{split}
    F&=\dfrac{a}{2}(q_1^2+q_2^2)+\dfrac{b}{4}(q_1^2+q_2^2)^2+\dfrac{d}{2}q_1^2q_2^2 \\
    &+\dfrac{s}{2}\left[(\nabla q_1)^2+(\nabla q_2)^2\right] \quad .
\label{eq:Landau_Ginzburg_simple}
\end{split}
\end{equation}
 
\subsubsection{N\'{e}el wall}
For ease of calculation, we parametrize $q_1$ and $q_2$ with polar coordinates $\{Q, \phi\}$. For a fixed amplitude $Q_0$, we construct the Euler-Lagrange equation (\ref{eq:Landau_Ginzburg_eff}) with respect to $\phi(z)$ and obtain the following equation:

\begin{equation}
    \nabla^2\phi(z) = \dfrac{dQ_0^2}{4s} \text{sin}[4\phi(z)].
    \label{eq:Sine_Gordon}
\end{equation}

Using the boundary conditions $Q(-\infty)=-Q_0$ and $Q(\infty)=Q_0$, equation (\ref{eq:Sine_Gordon}) is solved by a stationary Sine-Gordon equation and its solution is given by

\begin{align}
    \phi(z) &=  \arctan\left[\exp(z/\xi)\right] \label{eq:phi_analytical}, \\
    \xi &= \sqrt{\dfrac{s}{dQ_0^2}} \quad ,
    \label{eq:xi}
\end{align}

with $\xi$ considered to be half the domain wall width.

\subsubsection{Ising wall}
The other possible domain wall for a 2D order parameter is the Ising-type wall, in which the order parameter amplitude vanishes in the middle of the wall. (Note that for 1D irreps, this is the only possibility.) For such a wall, the Euler-Lagrange equation has the form 

\begin{equation}
    \nabla^2Q(z) = \dfrac{a}{2}Q(z) + \dfrac{b}{4}Q(z)^3\label{eq:isingdw},
\end{equation}
where the 6th-order Landau term has been omitted, so that we can exploit the known solutions of the 4rd-order equation. These solutions are 

\begin{align}
    Q(z) &=  Q_{0}\text{tanh}\left[\dfrac{z}{\sqrt{2}\xi}\right] \label{eq:Q_anlaytical}, \\
    Q_{0} &= \sqrt{-\dfrac{a}{b}} \label{eq:Q0}, \\
    \xi &= \sqrt{-\dfrac{s}{a}} \label{eq:xi2}.
\end{align}

\subsubsection{General domain wall profile}
In reality, a structural domain wall with 2D order parameter will be a mixture of N\'{e}el- and Ising types, and its profile can be obtained by solving both \eqref{eq:Sine_Gordon} and \eqref{eq:isingdw} simultaneously. This problem can likely not be solved analytically.

\subsection{Determination of the gradient parameter}\label{subsec:ginzburg_determination}
To calculate the gradient parameters we followed the procedure outlined by Artyukhin \textit{et al} in Ref.~\onlinecite{Artyukhin2013}. Consider any order parameter  $\vec{Q}$ described by the eigendisplacement $\vec{\eta}_Q(\vec{q})$ of a force constant mode corresponding to a wave vector $\vec{q}$. $\vec{Q}$ is then given by an eigenvector of the force constant matrix
\begin{equation}
    \vec{Q} \left(\vec{r} \right) = \vec{\eta}_{Q}(\vec{q}) e^{i\vec{q}\vec{r}}.
    \label{eq:displacements}
\end{equation}

The gradient energy term in $\vec{q}$-space associated with the parameter $s_Q$ can then be written as

\begin{equation}
    f_{G}(\vec{q}) = s_{Q} \left( \nabla \vec{Q} \right)^2 = s_{Q} \vec{q}^{\,2} |\vec{\eta}_{Q}|^2  \quad ,
    \label{Ginzburg}
\end{equation}
which equals 0 if $\vec{q}=0$. Therefore it is possible to determine $s_{Q}$ for some direction of $\vec{q}$ by calculating the energies of supercells with distortions described by Eqn.~\ref{eq:displacements} with various magnitudes of $\vec{q}$s frozen in along this direction. 

A more feasible approach is to determine $s_{Q}$ from the force constant dispersion. The Hessian of the gradient energy for all modes $\vec{\eta}_{q}$ of $\vec{q}$ is equal to the force constant matrix in $\vec{q}$-space $C\left( \vec{q}\right)$ within the harmonic approximation:

\begin{equation}
    \dfrac{\partial^2 f_G}{\partial \vec{\eta}^2_{q}} \left( \vec{q} \right) = s_{q} \vec{q}^{\,2} = C\left(\vec{q}\right).
    \label{eq:Hessian}
\end{equation}

Consequently, the eigenvalues of the Hessian in expression (\ref{eq:Hessian}) are the eigenvalues of the force constant matrix and we can determine all gradient parameters of modes with wave vector $\vec{q}$ by expanding the Hessian in (\ref{eq:Hessian}) around $\vec{q}=0$:

\begin{equation}
    s_{q} = \dfrac{1}{2} \dfrac{\partial^2 C(\vec{q})}{\partial \vec{q}^{\,2}}\biggr\rvert_{q=0}.
    \label{eq:sq}
\end{equation}

The gradient parameter corresponding to the mode $\eta_Q$ is then given by the eigenvalue of expression (\ref{eq:sq}) corresponding to the mode $\eta_Q$:

\begin{equation}
    s_Q = \left< \vec{\eta}_{Q} \rvert  s_q \rvert \vec{\eta}_{Q} \right>.
\end{equation}

\subsection{Exchange-correlation-functional suitability for the description of the $\beta$ to $\gamma$ transition} \label{subsec:gamma_domain_wall}

\begin{figure*}[t]
  \begin{minipage}[c]{\columnwidth}
    \null
\begin{tikzpicture}
    \node at (0,0)
        {\includegraphics[width=2\textwidth, trim={0.8cm 13.5cm 0.8cm 14cm},clip]{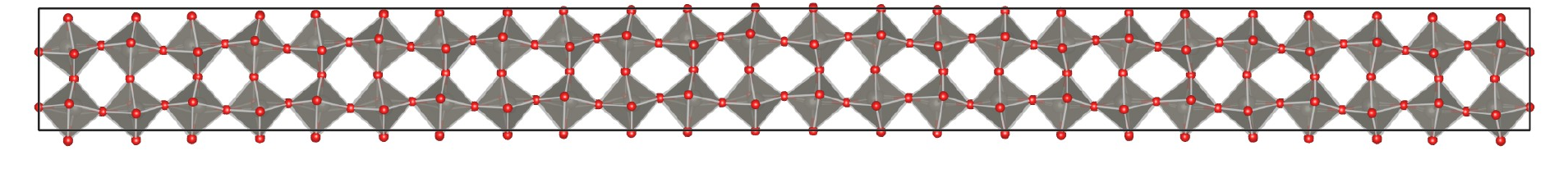}};
    \draw[very thick, blue] (-8.54,-0.63) coordinate (A) -- (-7.1,-0.59) coordinate (B) -- (B |- ,0.86) coordinate (C)
    -- (A |- ,0.82) -- cycle;
    \def \l {1.42}
    \draw[very thick, red] (-4.95,-0.5) -- ++(0:\l) --++(90:\l) --++(180:\l) --++(270:\l) -- cycle;
    \draw[very thick, red] (4.95,-0.5) -- ++(270:-\l) --++(180:\l) --++(90:-\l) --++(0:\l) -- cycle;
    \draw[very thick, black, dashed] (0,-0.8) --(0,1.2);
    \draw (0,-1.2)  node[below=2] {$M_2^+$, $X_5^+$} circle (0.16);
    \fill (0,-1.2)  circle (.04);
    \draw[very thick, ->] (-5,-1.2) node[below] {$M_2^-$} -- (-5,1.6);
    \draw[very thick, ->] (5,-1.2) node[below] {$M_2^-$} -- (5,1.6);
     \draw[very thick, black!30!green, ->] (-6,-1.2) node[below] {$X_5^+, R_5^-$} -- (-6,1.6);
    \draw[very thick, black!30!green, ->] (6,-1.2) node[below] {$X_5^+, R_5^-$} -- (6,1.6); 
    \draw[very thick, ->] (-3.5,1.5) -- ++(3:3);
    \draw[very thick, black!30!green, ->] (-3.2,1.5) arc (0:320:0.3);
    \draw node[above] at (-2,1.6) {$M_2^-, R_5^-$};
    \draw[very thick, ->] (3.5,1.5) -- ++(177:3);
    \draw[very thick, black!30!green, <-] (3.3,1.3) arc (-140:180:0.3);
    \draw node[above] at (2,1.6) {$M_2^-, R_5^-$};
\end{tikzpicture}
  \end{minipage}\hfill
  \begin{minipage}[c]{\columnwidth}

  \end{minipage}
   \caption{\protect Initial unrelaxed supercell for the $\gamma$ domain wall corresponding to a crystallographic $\left(100\right)$ plane. The monoclinic $\gamma$ unit cell is indicated by the blue box and the ions that were kept fixed during the relaxation by red boxes. The black vectors indicate the cubic phonon modes and directions that are already present in the $\beta$ phase; those introduced by the $\beta-\gamma$ transition are shown by green vectors. The orthorhombic $\Gamma_2^+$ mode causes a slight monoclinic tilt in one of the directions of the cubic $M_2^-$ and $R_5^-$ modes each of which is illustrated by circular arrows.}
    \label{fig:gamma_wall}
\end{figure*}
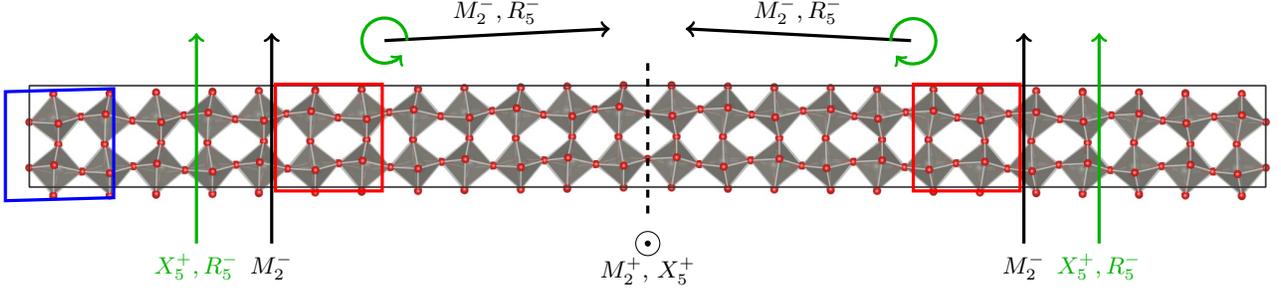

As mentioned in the main text, both LDA and GGA (PBEsol) exchange-correlation functionals yielded almost identical energies for the $\beta$ and $\gamma$ phases, even though the structural relaxations yielded distinct structures. This was the case for both {\sc Quantum Espresso} and {\sc VASP} calculations. 
Consequences of the small energy difference between the two phases were a negligible Landau parameter $a$, which (consistent with Eqn.~\ref{eq:xi2}) led to unreasonably large widths for the $\gamma$ domain walls, and erratic behavior in the DFT structural relaxations.

A crude estimate for the Landau parameters can be made by using the $\beta-\gamma$ transition energy obtained with B1-WC calculations as reported by Hamdi \textit{et al.} to approximate the Landau parameters $a$ and $b$ \cite{Hamdi2016}. The condition that the mode amplitude $Q_0$ in (\ref{eq:Q0}) equals the bulk amplitude of the orthorhombic $\Gamma_2^+$ mode in the $\gamma$ phase (experimentally \SI{0.531}{\angstrom}) and that the Landau potential in eq. (\ref{eq:Landau_Ginzburg_eff}) with only the $a_Q$ and $b_Q$ terms equals this reported energy difference ($F_{Pbcn}^0$-\SI{13}{\milli\electronvolt}/f.u.) at $Q_0$ results in the values for $a_Q$ and $b_Q$ of \SI{-0.22}{\milli\electronvolt\per\angstrom\tothe{5}} and \SI{0.39}{\milli\electronvolt\per\angstrom\tothe{7}} which are comparable to our values for the $\alpha_2-\beta$ transition Landau potential \cite{Vogt1999, Hamdi2016}. However, ultimately a final estimation for the Ising domain wall width cannot be made without the Ginzburg parameter $s$ and the calculation thereof for the orthorhombic $\Gamma_2^+$ mode does not seem reasonable based upon the poor characterization of the $\gamma$ phase by LDA.

\subsection{Superconducting critical temperature}\label{subsec:superconductivity}
Once the electron-phonon matrix in equation (\ref{eq:ep_matrix}) is known, the phonon linewidth $\gamma_{\vec{q}\nu}$ resulting from electron-phonon interaction can be calculated \cite{Migdal1958, McMillan1968, Allen1975, Giustino2007, Noffsinger2010, Ponce2016}. Within the Migdal approximation the linewidth of a phonon with wave vector $\vec{q}$ and branch $\nu$ is given as the imaginary component of the phonon self-energy,
\begin{multline}
   \gamma_{\vec{q}}^{\nu} = \text{Im} \sum_{nm} \dfrac{1}{\Omega_{BZ}}\int_{BZ} w_{\vec{k}} \lvert g_{nm}^{\nu} \left( \vec{k}, \vec{q} \right) \rvert^2 \\  \times \dfrac{f(\epsilon_{n\vec{k}})-f(\epsilon_{m\vec{k}+\vec{q}})}{\epsilon_{n\vec{k}}-\epsilon_{m\vec{k}+\vec{q}} - \omega_{\vec{q}}^{\nu}+ i\eta} d\vec{k},
   \label{eq:phonon_linewidth}
\end{multline}
   where $w_k$ denotes the weights for the $k$-points, $\epsilon_{n\vec{k}}$ is the band energy, $f(\epsilon)$ is the associated Fermi occupancy, $\omega_{\vec{q}}^{\nu}$ is the phonon frequency and $\eta$ is a smearing parameter for allowed transitions. In principle, the latter can be neglected in calculations where the $k$- and $q$-grids are dense enough. In such a limit of vanishing smearing, and additionally vanishing phonon frequencies, $\lim_{\eta, \omega_{\vec{q}\nu}\to0}  \gamma_{\vec{q}\nu}$, one arrives at the so-called double-delta approximation of the phonon linewidth

\begin{equation}
	\gamma_{\vec{q}}^{\nu} = 2\pi \omega_{\vec{q}}^{\nu} \sum_{nm} \dfrac{1}{\Omega_{BZ}}\int_{BZ} d\vec{k} w_{\vec{k}} |g_{nm}^{\nu} (\vec{k},\vec{q}) | ^2 \delta(\epsilon_{n\vec{k}}) \delta(\epsilon_{m\vec{k}+\vec{q}}),
	\label{eq:double_delta}
\end{equation}

where a smearing may be reintroduced in the the two delta functions. A similar expression as the one for the double-delta approximation (\ref{eq:double_delta}) gives the electron-phonon coupling strength $\lambda_{\vec{q}}^{\nu}$ for the phonon 

\begin{equation}
	\lambda_{\vec{q}}^{\nu} = \dfrac{1}{n(E_F)\omega_{\vec{q}}^{\nu}} \dfrac{1}{\Omega_{BZ}}\int_{BZ} d\vec{k} |g_{nm}^{\nu} (\vec{k},\vec{q}) | ^2 \delta(\epsilon_{n\vec{k}}) \delta(\epsilon_{m\vec{k}+\vec{q}}),
	\label{eq:coupling_strength}
\end{equation}
where $n(E_F)$ is the DOS at the Fermi energy.
Consequently, the coupling strength in the double-delta approximation is given as
\begin{equation}
	\lambda_{\vec{q}}^{\nu} = \dfrac{\gamma_{\vec{q}}^{\nu}}{\pi n(E_F){\omega_{\vec{q}}^{\nu}}^2}.
\end{equation}

Calculating the Brillouin-zone average of the coupling strength yields the first parameter in the McMillan formula which is the total coupling strength $\lambda$

\begin{equation}
	\lambda = \sum_{\nu} \dfrac{1}{\Omega_{BZ}}\int_{BZ} d\vec{q} w_{\vec{q}} \lambda_{\vec{q}}^{\nu},
\end{equation}

where $w_{\vec{q}}$ now denotes the weights for the $q$ points. In the Allen-Dynes formula, $\lambda$ is evaluated as 

\begin{equation}
	\lambda = 2 \int \dfrac{d\omega}{\omega} \alpha^2 F (\omega),
	\label{eq:lambda}
\end{equation}

where $\alpha^2 F(\omega)$ is the isotropic Eliashberg spectral function, which in turn is given as 

\begin{equation}
	\alpha^2 F(\omega) = \dfrac{1}{2}\sum_{\nu} \dfrac{1}{\Omega_{BZ}}\int_{BZ} d\vec{q} w_{\vec{q}} \omega_{\vec{q}}^{\nu}\lambda_{\vec{q}}^{\nu}\delta (\omega - \omega_{\vec{q}}^{\nu}).
	\label{eq:eliashberg}
\end{equation}

The delta function may again be subject to a smearing for numerical calculations. The remaining characteristic phonon frequency $\left<\omega\right>$, according to Allen and Dynes, is then given as

\begin{equation}
    \left<\omega\right> = \text{exp}\left[\dfrac{2}{\lambda} \int \dfrac{d\omega}{\omega}\alpha^2F(\omega)\text{log}(\omega)\right].
    \label{eq:freq}
\end{equation}
\newpage
\subsection{Calculated Landau-Ginzburg parameters}
\renewcommand{\arraystretch}{2}
\setlength{\tabcolsep}{10pt} 
\begin{table}[h!]
    \centering
    \resizebox{0.8\columnwidth}{!}{%
    \begin{tabular}{l|c}
         \hline
         \hline
         $\mu \left[meV\mathring{A}^{-3}\right]$ & 0.17 $\rho^2$\\
         $s$ & 6.64 + 4.59$\rho^2$ + 1.68$\rho^4$\\
         $t$ & 1026.92 + 70.9$(\rho+2.41)^2$ \\
          & + 16.6$(\rho+2.41)^4$\\

         \hline
         
         $a_R \left[meV\mathring{A}^{-4}\right]$ & $0.99\rho+0.04\rho^2$\\

         \hline
         
         $a_Q  \left[meV\mathring{A}^{-5}\right]$ & $-0.30 -0.59\rho -0.04\rho^2$\\
         $b_R$ & $1.59 - 0.49\rho -0.48\rho^2$\\

         \hline
        
         $c_R \left[meV\mathring{A}^{-6}\right]$ & $-0.95+0.5\rho+1.22\rho^2$ \\
         $a_{RQ}$ & $-0.85 -0.25\rho - 0.12 \rho^2$ \\

         \hline
         
         $b_Q \left[meV\mathring{A}^{-7}\right]$ & $0.26 +0.02\rho -0.008\rho^2$ \\
         $d_R$ & $0.28-0.74\rho - 1.67\rho^2$ \\
         $b_{RQ}$ & $0.34 + 0.52\rho + 0.26\rho^2$\\
         $a_{2Q}$ & $1.44 - 0.009\rho - 0.05 \rho^2$\\
         
         \hline
         
         $e_R \left[meV\mathring{A}^{-8}\right]$ & $-0.05 + 0.74\rho + 1.18\rho^2$ \\
         $c_{RQ}$ & $-0.22 -0.33\rho - 0.2 \rho^2$ \\
         $d_{RQ}$ & $0.08 + 0.07\rho + 0.04 \rho^2$\\
         $b_{2Q}$ & $-0.31 + 0.11\rho + 0.06\rho^2$\\
         $a_{R2Q}$ & $-0.004 + 0.003\rho + 0.002\rho^2$ \\

         \hline
         
         $c_Q \left[meV\mathring{A}^{-9}\right]$ & $10^{-6} + 0.009\rho + 0.002\rho^2$\\
         $f_R$ & $10^{-6} - 0.27\rho -0.33\rho^2$ \\
         $e_{RQ}$ & $0.11 + 0.05\rho + 0.05 \rho^2$\\
         $f_{RQ}$ & $2*10^{-6} - 0.03\rho -0.03 \rho^2$\\
         $c_{2Q}$ & $0.04 - 0.14\rho - 0.11\rho^2$ \\
         $d_{2Q}$ & $0.02 - 0.03\rho + 0.02 \rho^2$ \\
         $b_{R2Q}$ & $0.24 + 1.46\rho + 0.72\rho^2$\\ 

         \hline
         
         $c_{R2Q} \left[meV\mathring{A}^{-10}\right]$ & $-0.75 - 1.81\rho - 1.03 \rho^2$\\
         $d_{R2Q}$ & $-0.09 - 0.13\rho - 0.11\rho^2$ \\
         
         \hline
         
         $e_{R2Q} \left[meV\mathring{A}^{-11}\right]$ & $0.41 + 0.76\rho + 0.47\rho^2$\\
         $f_{R2Q}$ & $10^{-6} - 0.25\rho - 0.07\rho^2$\\
         
    \end{tabular}
    }
    \caption{Landau-Ginzburg parameters and their dependence on doping calculated to sixth order using DFT for a doping range of -0.25 to \SI{0}{\milli\elementarycharge\per\cubic\angstrom}.}
    
    \label{tab:Landau_parameters}

\end{table}

\subsection{Auxiliary plots}

\begin{figure}[!h]
    \centering
    \includegraphics[width=0.35\textwidth]{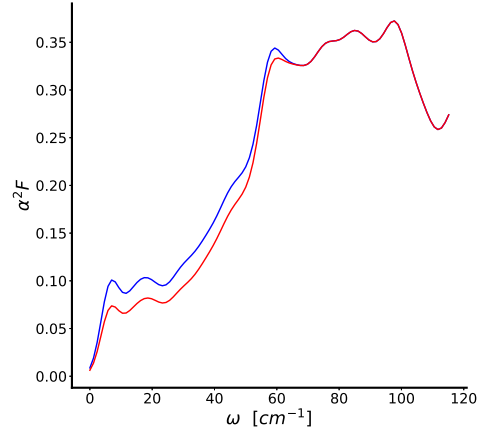}
    \caption{Eliashberg spectral function $\alpha^2F$ as a function of frequency in the low-frequency range for the $\alpha_1$ phase, at a doping level of $x=0.18$ (blue curve), and modified by excluding electron-phonon couplings around $\omega=0, q=[1/2, 1/2, 0]$ (red curve). The full range is shown in Fig.~\ref{fig:ph_e} }
    \label{fig:a2F_change}
\end{figure}

\begin{figure}[h]
    \centering
    \includegraphics[width=0.4\textwidth]{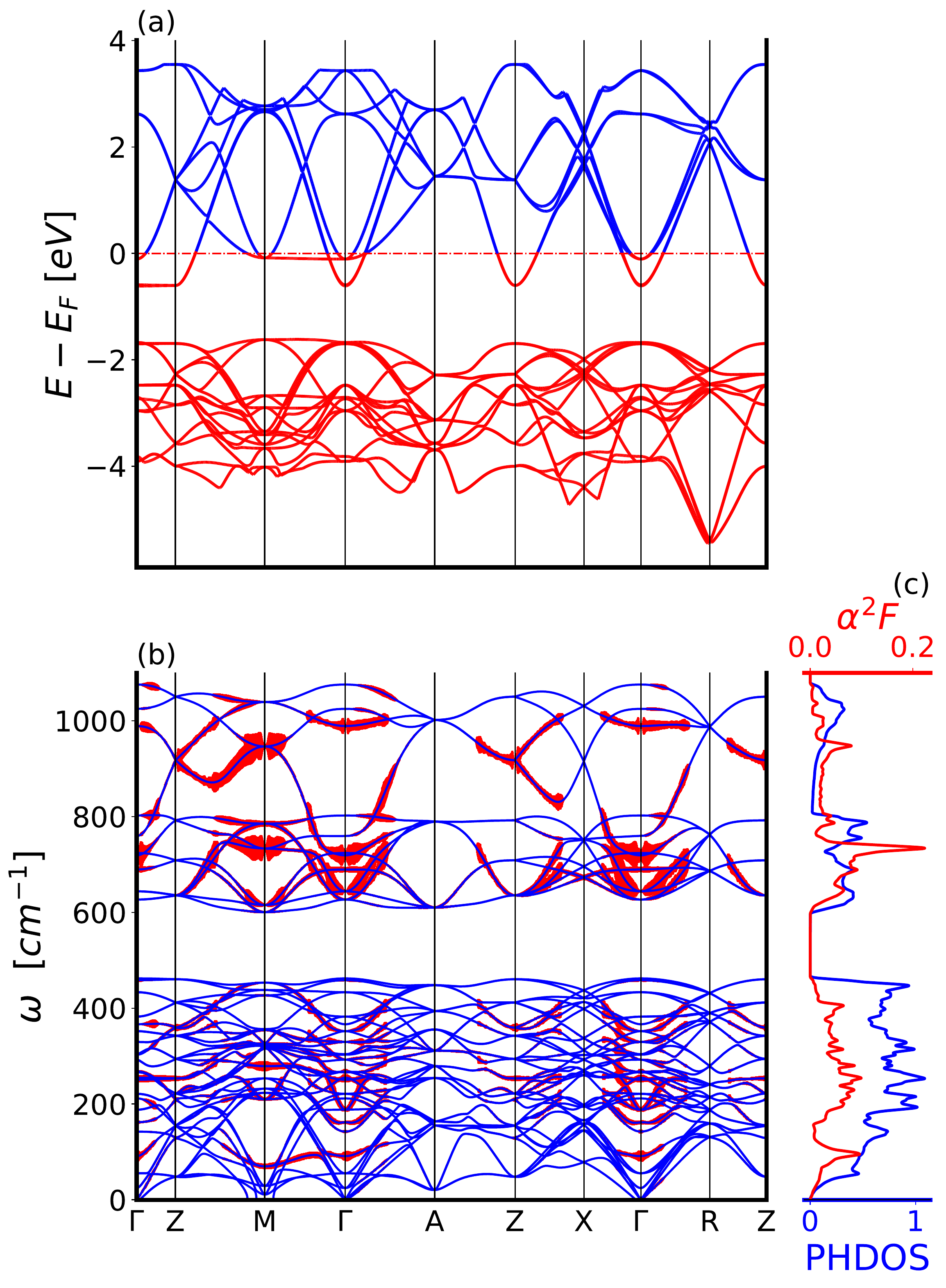}
    \caption{Calculated electronic bands (a), phonon bands and phonon linewidths (b), phonon density of states and $\alpha^2F$ (c) for $x=0.125$ additional electrons per f.u. in the $\alpha_2$ phase. The phonon linewidths are shown as vertical red bars in the phonon plots and they are scaled by a factor of 10 for visibility. Occupied states in the electron band structure are shown in red, unoccupied states in blue. Horizontal line at zero  denotes the Fermi level.}
    \label{fig:phlw2}
\end{figure}
\clearpage

\bibliography{references}

\end{document}